\documentclass[12pt]{iopart}
\usepackage{graphicx,psfrag}
\usepackage{color}
\usepackage{xspace}
\usepackage{iopams,bbm}
\usepackage{bm}
\usepackage{psfrag}

\def\op#1{\hat{#1}}                               
\newcommand{\unitmatrix}{\mathbb{I}}
\newcommand{\gopp}[2]{\hat{\tilde{#1}}_{#2} \hspace*{-0.8em}{\phantom{#1}'^{\dagger}}}
\newcommand{\ket}[1]{{| #1 \rangle}\xspace}
\newcommand{\rket}[1]{{| #1 \rangle}_R\xspace}        
\newcommand{\bra}[1]{\langle #1 |\xspace}
\newcommand{\lbra}[1]{\xspace \hspace*{-0.5em}{\phantom{\ket{}}}_L \langle #1 | \xspace}
\newcommand{\braket}[2]{{\langle #1| #2 \rangle}\xspace}
\newcommand{\lrbraket}[2]{\hspace*{-0.5em}{\phantom{\ket{}}}_L{\langle #1| #2 \rangle}_R \xspace}
\newcommand{\aver}[3]{{\langle #1| #2 |#3 \rangle}\xspace}

\begin{document}

\title[Phaseless quantum Monte-Carlo approach]{Phaseless quantum Monte-Carlo
  approach to strongly correlated superconductors with
  stochastic Hartree-Fock-Bogoliubov wavefunctions} 
\author{Olivier Juillet$^{1}$, Alexandre Lepr\'evost$^{1,2}$, J\'er\'emy
  Bonnard$^{1,3}$, and Raymond Fr\'esard$^{2}$}
\address{$^1$ Normandie Universit\'e, ENSICAEN, UNICAEN, CNRS, LPC, 14050 Caen,
  France}
\address{$^2$ Normandie Universit\'e, ENSICAEN, UNICAEN, CNRS, CRISMAT, 14050
  Caen, France}
\address{$^3$ Institut de Physique Nucl\'eaire, CNRS-IN2P3, Universit\'e
  Paris-Sud, Universit\'e Paris-Saclay, 91406 Orsay Cedex, France} 

\begin{abstract}
The so-called phaseless quantum Monte-Carlo method currently offers one of the
best performing theoretical framework to investigate interacting Fermi
systems. It allows to extract an approximate ground-state wavefunction by
averaging independent-particle states undergoing a Brownian motion in
imaginary-time. Here, we extend the approach to a random walk in the space of
Hartree-Fock-Bogoliubov (HFB) vacua that are better suited for superconducting
or superfluid systems. Well-controlled statistical errors are ensured by
constraining stochastic paths with the help of a trial wavefunction. It also
guides the dynamics and takes the form of a linear combination of HFB
ans\"atze. Estimates for the observables are reconstructed through an
extension of Wick's theorem to matrix elements between HFB product states. The
usual combinatory complexity associated to the application of this theorem for
four- and more- body operators is bypassed with a compact expression in terms
of Pfaffians. The limiting case of a stochastic motion within Slater
determinants but guided with HFB trial wavefunctions is also
considered. Finally, exploratory results for the spin-polarized Hubbard model
in the attractive regime are presented.  
\end{abstract}
\maketitle

%

\section{\label{sec:introduction}Introduction}

Frictionless flow is one of the most spectacular manifestation of quantum
coherence in many-body systems at the macroscopic scale. Historically, its
appearance in fermion matter has been first evidenced in superconducting (SC)
metals, nuclei and superfluid $^3$He. Currently, Cooper-pair condensates
keep on attracting interest, especially because of the diversity of observed
ground states. For a large class of SC materials, ranging from elements (such
as Hg \cite{Kam11}) to alloys (such as Nb$_3$Ge \cite{Mat65}), and possibly to
the newly discovered high-$T_c$ hydrogen sulfide under pressure \cite{Dro15},
the pair wavefunction exhibits $s$-wave symmetry. The microscopic mechanism for
electron pairing is 
then well established and invokes a phonon mediated attraction according to
Bardeen-Cooper-Schrieffer (BCS) and Eliashberg theories. However, condensed
matter physics also harbors an ever increasing family of superconductors
that challenges this conventional paradigm. This applies in particular to heavy
fermion systems, where the formation of local Cooper pairs is suppressed by
the Coulomb interaction. 
As reviewed by, e.g., Thalmeier et al. \cite{RevSt}, this leads to a wealth
of behaviors. For example,  SC pairing due to magnetic excitons has been
observed in UPd$_2$Al$_3$ \cite{Jour}. Alternatively, comprehensive
experimental work demonstrated that in several Ce-based systems, including
CeCu$_2$Si$_2$, 
SC tends to mostly appear in the vicinity of a quantum critical point
\cite{Yuan03,Bruls94}, thereby providing strong evidence of the 
interplay between magnetic and pairing degrees of freedom. 

Strong correlation is also a hallmark of the superconducting cuprates, where
numerous experiments point towards an SC order parameter with $d$-wave symmetry.
Furthermore, it has long been suspected on theoretical grounds, that
unconventional SC pairing mechanism involving momentum-carrying Cooper pairs
is favored in striped phases \cite{Him02,Rac07}. The recent observation of
pair-density waves in Bi$_2$Sr$_2$CaCu$_2$O$_{8+x}$ \cite{Ham16} came
to support this scenario, that strongly suggests an intertwining of the spin,
charge, and pair degrees of freedom \cite{Fra15,Lep15}. More generally, the SC
correlations in cuprates develop from a bad metal normal phase and in the
vicinity of an antiferromagnetic order. A similar feature is in fact shared by
other classes of high-$T_c$ SC compounds, such as iron pnictides and
chalcogenides \cite{Pnic08}. These materials are nevertheless singular by
exhibiting 
multi-band Fermi surfaces, leading to a non-universal and still under debate
SC order parameter where extended $s$-wave and $d$-wave pairing symmetries are
close competitors. Finally, ruthenate superconductors currently attract a
particular attention due to a spin-triplet $p$-wave pair condensation, possibly
induced by ferromagnetic spin fluctuations. In addition, a
variety of experiments point toward a chiral SC order parameter that could
support non-Abelian excitations \cite{Mae01}. 

Understanding most of the above mentioned strongly correlated  SC systems
obviously requires going beyond the standard BCS mean-field approximation that
assumes independent Bogoliubov quasiparticles. For lattice electron models,
such as the single- or multi-band Hubbard Hamiltonians, Gutzwiller-BCS
wavefunctions are thus frequently used as trial states. In this case, strong
electronic correlations are included by partially or totally suppressing
double occupancy entailed in a BCS ansatz with an assumed internal structure
of the fermion pairs \cite{Gia91, Mis14}. Alternatively, the variational optimization of large
coherent superpositions of general BCS states, free of any \textit{a priori} input on
the relevant correlations, have been recently reported \cite{Lep15}. However, Quantum
Monte-Carlo (QMC) methods remain indubitably the most powerful approaches to
solve the Schr\"odinger equation in many-body systems. The acronym QMC actually
embraces a multitude of stochastic algorithms, but the general strategy is to
represent the zero- or finite-temperature equilibrium state as an integral in
a high-dimensional space that can be evaluated using random walks. A
compelling example is the auxiliary-field QMC approach that performs
projection on exact ground states by sampling fictitious systems of
independent particles in external fluctuating fields. The half-filled Hubbard
model on bipartite lattices has been extensively investigated through such
techniques \cite{Hir85, Whi89}. For instance, the Mott metal-to-insulator transition on the
honeycomb lattice was recently addressed on clusters which size exceeds 2000
sites \cite{Men10,Sor12}. Another investigated issue is the entanglement of
Mott insulators through the determination of the Renyi entropies \cite{Gro13, Ass14}.

Unfortunately, the general applicability of fermionic QMC schemes suffers from
the emergence of negative ``weights'' or even complex contributions, that
deteriorate the stochastic reconstruction of the exact wavefunction to a point
where the signal-to-noise ratio becomes almost zero. This infamous sign or
phase problem has been proven to be NP-hard and its algorithmic solution with
polynomial-time complexity probably does not exist \cite{Tro05}. For QMC
methods in the configuration space \cite{Cep80, Rey82, Foul01} and
time-invariant Hamiltonians, a solution could be found in principle by
preventing the random walk to move across a node of the exact ground state. In
practice, the nodal surface is imposed from a trial wavefunction and
calculations become variational. With auxiliary-field QMC approaches,
sufficient conditions for the absence of sign/phase problems are known
\cite{Wu05}. Otherwise, restricted-path approximations have been developed for 
lattice models with Hubbard-like Hamiltonians that also ensure a sampling of
real Slater determinants \cite{Zha95, Zha99}. The so-called phaseless QMC
scheme developed by Zhang and Krakauer \cite{Zha03} 
finally provides the most general framework to obtain an approximate
ground state of any fermion system defined either by a model or by a realistic
Hamiltonian 
including two-body interactions. In the original formulation, the 
method relies on a random walk within independent-particle states that is
guided and constrained to control phase problems thanks to a trial many-body
wavefunction. Up to now, single- or multi-determinant states have been used
for this purpose. 

In the present paper, we propose an extension better suited for
superconductors through BCS wavefunctions not only as trial states, but also
as walkers to absorb in a single path fermion-pair condensation. We first
review in Section~\ref{sec:HFB} the basic properties of the most general pair
coherent states, known as Hartree-Fock-Bogoliubov (HFB) wavefunctions and for
which relevant Cooper pairs do not need to be assumed. The imaginary-time
many-body Schr\"odinger equation is then reformulated in Section~\ref{sec:sto}
as the average of stochastic trajectories of such HFB states. The
restricted-path approximations required to set up a viable QMC scheme,
i.e. free of phase problems and with an ensured finite-variance sampling, are
presented in Section~\ref{sec:variance}. The estimation of ground-state observables
is discussed in Section~\ref{sec:estobs} by extending the usual Wick theorem to
matrix elements between HFB states. Section~\ref{sec:numex} aims at illustrating
the approach and provides a first numerical implementation for a spin-polarized Fermi system in the 
strongly attractive regime. The last section is
devoted to conclusions and perspectives.  

All along this work, we consider a general system of fermions interacting \textit{via} a
two-body potential. The associated Hamiltonian will therefore be given, in
second quantization, by 
\begin{equation}\label{EQ:1}
\op{H} = \sum_{i, j} h_{i, j} \op{c}^{\dagger}_{i} \op{c}^{\vphantom{dagger}}_{j} +
\frac12 \sum_{i, j, k, l} V_{i, j, k, l}
\op{c}^{\dagger}_{i} \op{c}^{\dagger}_{j}
\op{c}^{\vphantom{dagger}}_{l} \op{c}^{\vphantom{dagger}}_{k}\,.
\end{equation}
$\op{c}^{\dagger}_{i} $ and $\op{c}^{\vphantom{dagger}}_{i}$ are the usual
creation and annihilation operators of a fermion in the state $\ket{i}$
belonging 
to an orthonormal basis in the one-body space $\mathcal{H}^{(1)}$, which is
supposed to be of finite size $d$.
\begin{eqnarray}
h_{i, j} = \aver{i}{\op{h}}{j}, \nonumber \\
V_{i, j, k, l} = \aver{i}{\otimes  \aver{j}{\op{V}}{k} \otimes}{l},
\end{eqnarray}
 represent respectively the matrix elements
of the one-body Hamiltonian and the binary interactions. In
  the context of a multiband Hubbard model, $i$ stands for a combined site,
  spin and orbital index. 

\section{The manifold of HFB wavefunctions} \label{sec:HFB}

All zero-temperature QMC methods rely on a numerical reconstruction of the
ground state using a stochastic exploration of a basis. For example, the
``Diffusion Monte-Carlo'' \cite{Cep80,Rey82,Foul01,And75} method uses the
orthonormal basis of the position representation, which is sampled by means
of a random walk or through a Brownian motion associated to the kinetic 
energy. Alternatively, auxiliary-field QMC approaches, which we will consider
in the following, traditionally call upon  the
overcomplete basis of independent-particle states for a system of $N$ fermions
\begin{equation}
\ket{\Phi} = \prod_{n=1}^{N} \op{c}^{\dagger}_{\phi_n}\ket{\,},\,\,\, \mbox{with} \,\,\,
\op{c}^{\dagger}_{\phi_n} = \sum_i \op{c}^{\dagger}_{i} \phi_{i,n} \,.
\end{equation}
Here, $\{\phi_{i,n}\}$ denote the components in the basis $\{\ket{i}\}$ of individual occupied and 
orthonormal states. In this work, we propose to extend the formalism to the most general factorized
fermionic states, i.e. Hartree-Fock-Bogoliubov (HFB) wavefunctions \cite{Bla85, Rin80} ---also 
known in condensed-matter Physics as Bogoliubov-de Gennes wavefunctions---,
\begin{equation}\label{EQ:2}
\ket{\Phi} \propto \prod_{n=1}^{d} \op{\gamma}_n \ket{\,} \,,
\end{equation}
where the quasiparticle operators $\op{\gamma}_n $ $( n \in \{1,\ldots ,d\})$
obey the Fermi-Dirac statistics and are linearly related to the original
creation/annihilation fermionic operators
\begin{equation}\label{eq:12}
\op{\gamma}_n = \sum_{i} 
\op{c}_{i}^{\dagger} V^*_{i,n} + 
\op{c}_{i}^{\vphantom{\dagger}} U^*_{i,n},
\end{equation}
that is 
\begin{equation}\label{eq:9p}
\left(\begin{array}{l}
\op{\gamma}\\
\op{\gamma}^{\dagger}
\end{array}\right) = \left(\begin{array}{cc}
U^{\dagger}&V^{\dagger}\\
V^{T}&U^{T}
\end{array}\right) 
\left(\begin{array}{l}
\op{c}\\
\op{c}^{\dagger}
\end{array}\right) \equiv \mathcal{B}^{\dagger} 
\left(\begin{array}{l}
\op{c}\\
\op{c}^{\dagger}
\end{array}\right).
\end{equation}
Thus, $\op{\gamma}_n \ket{\Phi} = 0$ and the HFB state appears as a quasiparticle vacuum. 
The Bogoliubov transformation matrix $\mathcal{B}$ (\ref{eq:9p}) will
moreover be supposed unitary to ensure canonical anticommutation relations 
$[\op{\gamma}^{\vphantom{\dagger}}_n ,\op{\gamma}^{\vphantom{\dagger}}_p]_+=0
$, $[\op{\gamma}^{\vphantom{\dagger}}_n ,\op{\gamma}^{\dagger}_p]_+=
\delta_{n,p} \op{\unitmatrix}$, and to allow the expansion of the physical operators in terms of 
quasiparticles
\begin{equation}\label{eq:9q}
\left(\begin{array}{l}
\op{c}\\
\op{c}^{\dagger}
\end{array}\right) = \left(\begin{array}{cc}
U&V^{*}\\
V&U^{*}
\end{array}\right) 
\left(\begin{array}{l}
\op{\gamma}\\
\op{\gamma}^{\dagger}
\end{array}\right) ,
\end{equation}
or, equivalently
\begin{equation}\label{eq:12p}
\op{c}_i = \sum_{n} 
\left( U^{\vphantom{*}}_{i,n} \op{\gamma}^{\vphantom{\dagger}}_n + 
V^*_{i,n} \op{\gamma}^{\dagger}_n \right) .
\end{equation}

Mathematically, the HFB wavefunctions (\ref{EQ:2}) form an overcomplete
set of the Fock space due to their quality of coherent states associated to
the Lie algebra $SO(2d)$ of operators 
$\op{c}^{\dagger}_i \op{c}^{\vphantom{\dagger}}_j -\frac12\delta_{i,j}$, 
$\op{c}^{\dagger}_i \op{c}^{\dagger}_j$, and 
$\op{c}^{\vphantom{\dagger}}_i \op{c}^{\vphantom{\dagger}}_j$
\cite{Zha90}. As a consequence, every correlated state of a fermionic system
can be reconstructed as a linear combination of non-orthogonal HFB
wavefunctions. By breaking the $U(1)$ gauge symmetry of the Hamiltonian (\ref{EQ:1}), HFB states 
physically present the advantage of immediately
leading to pairing correlations. A single HFB wavefunction may display the
generic properties of superconducting or superfluid fermionic phases, while
large coherent superpositions are required to reproduce such behaviors 
with independent-particle states. 

Historically, the condensation of Cooper pairs is however rather apprehended
through the BCS wavefunction expressed in terms of couples
$\left(\ket{\phi_{\alpha}}, \ket{\phi_{\tilde{\alpha}}}\right)$ of one-body
states between which pairing is established, and which form an orthonormal
basis of $\mathcal{H}^{(1)}$
\begin{equation}\label{EQ:5}
\ket{\Phi^{({\rm BCS})}} = \prod_{\alpha=1}^{d/2}\left( u_{\alpha} + v_{\alpha}
\op{c}^{\dagger}_{\phi_{\alpha}} \op{c}^{\dagger}_{\phi_{\tilde{\alpha}}} \right) \ket{\,}.
\end{equation}
$u_{\alpha}$ and $v_{\alpha}$ represent the probability amplitudes for the
couple $\left(\ket{\phi_{\alpha}}, \ket{\phi_{\tilde{\alpha}}}\right)$ to be
unoccupied or populated, respectively. In variational treatments, the BCS
mean-field method is based on the minimization of the energy in the state
(\ref{EQ:5}) with respect to the parameters $(u_{\alpha},v_{\alpha})$,
the pairs $\op{c}^{\dagger}_{\phi_{\alpha}}
\op{c}^{\dagger}_{\phi_{\tilde{\alpha}}}$ being chosen on physical
considerations. The HFB ansatz (\ref{EQ:2}) actually has the advantage of
completing the description by determining the optimal basis 
$\left(\ket{\phi_{\alpha}}, \ket{\phi_{\tilde{\alpha}}}\right)$. It may indeed
be re-written in a BCS form with states $\left(\ket{\phi_{\alpha}},
  \ket{\phi_{\tilde{\alpha}}}\right)$ directly encoded in the Bogoliubov
transformation. The evidence of this result is based on the
Bloch-Messiah-Zumino theorem \cite{Blo62, Zum62}, which states that there
exists unitary matrices $C$ and $D$  such that
\begin{eqnarray}\label{eq:15-15p}
D^{\dagger} U C^{\dagger} &= 
\left(\begin{array}{llll} 
\underline{U}_1&0_{2\times 2}& \ldots & 0_{2\times 2}\\
0_{2\times 2}&\underline{U}_2& \ldots & 0_{2\times 2}\\
\vdots&\vdots&\ddots &\vdots\\ 
0_{2\times 2}&0_{2\times 2}&\ldots &\underline{U}_{d/2}
\end{array}\right) \equiv \underline{U} , \nonumber \\
D^{T} V C^{\dagger} &=
\left(\begin{array}{llll} 
\underline{V}_1&0_{2\times 2}& \ldots & 0_{2\times 2}\\
0_{2\times 2}&\underline{V}_2& \ldots & 0_{2\times 2}\\
\vdots&\vdots&\ddots &\vdots\\ 
0_{2\times 2}&0_{2\times 2}&\ldots &\underline{V}_{d/2}
\end{array}\right) \equiv \underline{V},
\end{eqnarray}
where the $2\times 2$ diagonal blocks $\underline{U}_{\alpha}$ and
$\underline{V}_{\alpha}$ $(\alpha \in \{1,\ldots, d/2\})$ are respectively
characterized by two real positive numbers $u_{\alpha}, v_{\alpha}$, and are given by
\begin{equation}
\underline{U}_{\alpha} =
\left(\begin{array}{cc}
u_{\alpha}&0\\
0&u_{\alpha}
\end{array}\right), 
\quad
\underline{V}_{\alpha} =
\left(\begin{array}{cc}
0&v_{\alpha}\\
-v_{\alpha}&0
\end{array}\right) \,.
\end{equation}
One can then define the orthonormal basis $\left(\ket{\phi_{\alpha}},
  \ket{\phi_{\tilde{\alpha}}}\right)$ as the one obtained by applying the
unitary transformation $D$ on the family $\ket{i}$,
\begin{equation}
\ket{\phi_{\alpha}} = \sum_i \ket{i} D_{i, 2\alpha -1}\,,\quad
\ket{\phi_{\tilde{\alpha}}} = \sum_i \ket{i} D_{i, 2\alpha}.
\end{equation}
Through the previous decomposition of the matrices $U$ and $V$ in their
respective canonical form $\underline{U}$ and $\underline{V}$, the
quasiparticle operators $\op{\gamma}_n$ (\ref{eq:12}) of the HFB state can
be immediately obtained in terms of their counterparts
$\underline{\op{\gamma}}_{\alpha} = 
u_{\alpha} \op{c}^{\vphantom{\dagger}}_{\phi_{\alpha}} -
v_{\alpha} \op{c}^{\dagger}_{\phi_{\tilde{\alpha}}}$,
$\underline{\op{\gamma}}_{\tilde{\alpha}} =
u_{\alpha} \op{c}^{\vphantom{\dagger}}_{\phi_{\tilde{\alpha}}} +
v_{\alpha} \op{c}^{\dagger}_{\phi_{\alpha}} $ 
for the BCS wavefunction (\ref{EQ:5})
\begin{equation}
\op{\gamma}^{\vphantom{\dagger}}_n = \sum_{\alpha=1}^{d/2}\left( 
\underline{\op{\gamma}}^{\vphantom{\dagger}}_{\alpha} C^*_{2\alpha -1, n} +
\underline{\op{\gamma}}^{\vphantom{\dagger}}_{\tilde{\alpha}} C^*_{2\alpha, n}
\right)\,.
\end{equation}
Taking into account the commutative algebra of the different factors, one is
led to 
\begin{equation}
\prod_{n=1}^{d} \op{\gamma}_n \ket{\,} = \det\left( C^* \right)
\prod_{\alpha=1}^{d/2} \underline{\op{\gamma}}_{\alpha} \underline{\op{\tilde{\gamma}}}_{\alpha}\ket{\,}\,.
\end{equation}
In addition, all products
$\underline{\op{\gamma}}^{\vphantom{\dagger}}_{\alpha}
\underline{\op{\gamma}}^{\vphantom{\dagger}}_{\tilde{\alpha}}$ commute and the
action of each of them on the vacuum is equivalent to $
u_{\alpha} v_{\alpha} \op{c}^{\vphantom{\dagger}}_{\phi_{\alpha}}
\op{c}^{\dagger}_{\phi_{\tilde{\alpha}}}$ or 
$ u_{\alpha} v_{\alpha} \op{\unitmatrix} +
v_{\alpha}^2\op{c}^{\dagger}_{\phi_{\alpha}}\op{c}^{\dagger}_{\phi_{\tilde{\alpha}}}$. 
When setting 
\begin{equation}\label{EQ:10}
\ket{\Phi} = \frac{1}{\nu_1 \ldots \nu_{d/2}} \op{\gamma}_1 \ldots
\op{\gamma}_d \ket{\,} \,,
\end{equation}
the HFB state finally reduces, up to a phase factor $\det(C^*)$, to a
BCS ansatz with pairs directly stemming from the Bogoliubov transformation. 
From now on, we will adopt this form which clarifies the normalization
constant in the definition (\ref{EQ:2}). 

In the following, the manipulation of quasiparticle operators (\ref{eq:12})
will be greatly facilitated by adding a label $\varpi \in \{p,h\}$ to each individual
state, that indicates whether it refers to a particle ($p$) or to a hole ($h$). 
In other words, we define $\op{c}_{p i}^{\dagger} = \op{c}_{i}^{\dagger}$ and 
$\op{c}_{h i}^{\dagger} = \op{c}_{i}^{\vphantom{\dagger}}$. The extended
one-body space obtained this way identifies to the tensor product
$\mathcal{H}_{ex}^{(1)} = \mathcal{H}^{(ph)} \otimes \mathcal{H}^{(1)}$, where 
$\mathcal{H}^{(ph)}$ refers to the abstract space of dimension two underpinned by
an orthonormal basis $\{\ket{e_p}, \ket{e_h} \}$ related to the two flavors
``particle'', ``hole''. In this context, we can for example immediately check that
the unitary character of the Bogoliubov matrix is equivalent to providing the
extended space with an orthonormal basis built upon the vectors
\begin{equation}
\ket{\gamma_n} = \left( \begin{array}{l}
\ket{V^*_n}\\ \ket{U^*_n} \end{array}\right) ,\, 
\ket{\bar{\gamma}_n} = \left(\begin{array}{l}
\ket{U_n}\\ \ket{V_n} 
\end{array}\right),
\end{equation}
where $\ket{U_n} = \sum_{i} \ket{i} U_{i,n} $, $\ket{U^*_n} = \sum_{i}
\ket{i} U^*_{i,n} $ (as well as  $\ket{V_n}$ and $\ket{V^*_n}$)
are defined from the amplitudes of the transformation. Indeed,
\begin{equation}
\sum_{n=1}^d \left( \ket{\gamma_n} \bra{\gamma_n} + \ket{\bar{\gamma}_n}
  \bra{\bar{\gamma}_n} \right) = \unitmatrix \,, \,\,\, \left\{ \begin{array}{l} \braket{\gamma_n}{\gamma_p} = 
\braket{\bar{\gamma}_n}{\bar{\gamma}_p} = \delta_{n,p}\\
\braket{\gamma_n}{\bar{\gamma}_p} = 0 \end{array}
\right. \,.
\end{equation}
In addition, the relations (\ref{eq:9p}, \ref{eq:9q}) between the fermionic operators and the
quasiparticles are simply written as
\begin{equation}\label{EQ:13}
\left\{ \begin{array}{l} \op{\gamma}_n = \sum_{\varpi i} 
\op{c}_{\varpi i}^{\dagger} \braket{\varpi i}{\gamma_n} = \sum_{\varpi i} 
\braket{\bar{\gamma}_n}{\varpi i} \op{c}_{\varpi i}^{\vphantom{\dagger}}\\
\op{\gamma}_n^{\dagger} = \sum_{\varpi i} 
\op{c}_{\varpi i}^{\dagger} \braket{\varpi i}{\bar{\gamma}_n} = \sum_{\varpi i} 
\braket{\gamma_n}{\varpi i} \op{c}_{\varpi i}^{\vphantom{\dagger}} \end{array}
\right.,
\end{equation}
and
\begin{equation}\label{EQ:14}
\left\{ \begin{array}{l}
\op{c}_{\varpi i}^{\vphantom{\dagger}} = \sum_{n=1}^d \left( 
\braket{\varpi i}{\bar{\gamma}_n}\op{\gamma}_n + \braket{\varpi i}{\gamma_n}
\op{\gamma}_n^{\dagger} \right)\\
\op{c}_{\varpi i}^{\dagger} = \sum_{n=1}^d \left( 
\op{\gamma}_n \braket{\gamma_n}{\varpi i} + \op{\gamma}_n^{\dagger}
\braket{\bar{\gamma}_n}{\varpi i} \right) \end{array}
\right..
\end{equation}
These expansions show that the operator $\op{\gamma}_n$ depends linearly on
the ket $ \ket{\gamma_n}$ or on the bra $ \bra{\bar{\gamma}_n}$. Accordingly,
his adjoint $\op{\gamma}_n^{\dagger}$ is a linear function of the ket 
$\ket{\bar{\gamma}_n}$ or of the bra $ \bra{\gamma_n}$.
\section{Stochastic reformulation of the imaginary-time dependent 
Schr\"odinger equation with HFB walkers} \label{sec:sto}
Auxiliary-field QMC approaches are based on the dynamics in imaginary time of
an initial wavefunction $\ket{\Phi_0}$ in order to project it onto the ground
state $\ket{\Psi_G}$ of the Hamiltonian:
\begin{equation}\label{EQ:15}
\ket{\Psi_G} \propto \lim_{\tau \rightarrow \infty} \exp\bigl(-\tau \op{H}\bigr)
\ket{\Phi_0} .
\end{equation}
One typically chooses $\ket{\Phi_0}$ as an independent-particle state and,
thanks to the Hubbard-Stratonovich transformation \cite{Sug86, Hir83}, the 
$N$-body propagator $\exp(-\tau \op{H})$ is rewritten in the form of a
multidimensional integral of one-body propagators in fluctuating effective
potentials. For fermions, the exact ground state thus appears eventually as an
average of Slater determinants $\ket{\Phi_{\tau}}$, which individual states
follow a Brownian motion and are consequently called walkers
\begin{equation}\label{EQ:16}
\ket{\Psi_G} \propto \lim_{\tau \rightarrow \infty} \mathbb{E}[ \ket{\Phi_{\tau}}]\,,
\end{equation}
in which $\mathbb{E}[\ldots]$ stands for the ensemble average of a stochastic
process. Despite its use in most QMC simulations, such a scheme is not the
optimal reformulation of the Schr\"odinger equation in imaginary time for the
$N$-body state. Indeed, the broadening of the probability distribution of the
walkers, measured through the growth of the averaged quadratic distance
\begin{equation}
\mathbb{E}\Bigl[\bigl|\bigl|\rme^{-\tau \op{H}}\ket{\Phi_0} 
-\ket{\Phi_{\tau}}\bigr|\bigr|^2\Bigr],
\end{equation}
between the exact propagation and one of its stochastic realizations, is not
made minimal. A QMC scheme fulfilling such a criterion was identified in 2001
for a system of bosons \cite{Car01} and extended in 2002 to fermions
\cite{Jui03}. It is based on a one-body dynamics controlled by the
mean-field Hartree-Fock Hamiltonian and enhanced by one-particle one-hole stochastic
excitations. In addition, each realization is separated from the exact state
by a bounded distance so that the variance of any observable is guaranteed not
to diverge \cite{Jui03}. This method remains delicate to implement
numerically and suffers in general, as does the standard scheme, from the
usual sign/phase problem. It has essentially been applied to study the
development of pairing correlations in a one-dimensional system of cold atoms
trapped in a harmonic well or on a rotating torus  \cite{Jui04}.

In order to improve the efficiency of the standard auxiliary-field QMC
dynamics, Zhang \& Krakauer suggested to incorporate in the motion of
the walkers a complex importance function given by their overlap with a
previously chosen trial state $\ket{\Psi_T}$ not orthogonal to $\ket{\Psi_G}$ \cite{Zha03}
\begin{equation}\label{EQ:psig}
\ket{\Psi_G} \propto \lim_{\tau \rightarrow \infty} \mathbb{E}\left[\Pi_{\tau}
  \frac{\ket{\Phi_{\tau}}}{\braket{\Psi_T}{\Phi_{\tau}}}  \right].
\end{equation}
Here, the addition of the factor $\Pi$ is necessary for the stochastic scheme
to be equivalent to the Schr\"odinger equation in imaginary time, as we will see
below. To date, the main applications of the sampling (\ref{EQ:psig}) are
related to quantum chemistry with a trial state $\ket{\Psi_T}$ as well as
walkers $\ket{\Phi_{\tau}}$ being Slater determinants \cite{eeam}. Frustrated magnetic models have
also been apprehended with a similar random walk within matrix-product states
\cite{mps}. Finally, the approach has been generalized to rebuild, in the
context of nuclear structure, the ground state \cite{Bon13} as well as 
excited states \cite{Bon16} in each symmetry sector of the Hamiltonian \textit{via}
wavefunctions projected onto the related quantum numbers. In the following, we
extend the guided dynamics QMC scheme (\ref{EQ:psig}) to HFB-type walkers.

Let us first consider the exact propagation of the ansatz 
$\ket{W} = \Pi \ket{\Phi}/\braket{\Psi_T}{\Phi}$
built from a normalized Bogo\-liubov vacuum $\ket{\Phi}$ (\ref{EQ:10}), an
arbitrary test wavefunction $\ket{\Psi_T}$, and a
multiplicative complex variable $\Pi$. After an infinitesimal imaginary time
$\rmd\tau$, the Schr\"odinger equation leads to a correlated state, given to
first order in $\rmd\tau$, by
\begin{equation}\label{EQ:17}
\exp\bigl(-\rmd\tau\op{H}\bigr)\ket{W} =\ket{W} - \rmd\tau \op{H}\ket{W}.
\end{equation}

On the other hand, an elementary variation of $\ket{W}$ may be obtained by
slightly changing the HFB state by means of a transformation $\op{T} =
\exp\left(\rmd \op{x}\right)$, where $\rmd \op{x}$ is a general
one-body operator, with infinitesimal matrix elements 
and which possibly includes terms $\op{c}_{i}^{\dagger} \op{c}_{j}^{\dagger} $, 
$\op{c}_{i}^{\vphantom{\dagger}} \op{c}_{j}^{\vphantom{\dagger}}$ that break
particle-number conservation. With the notations of the extended space
$\mathcal{H}^{(1)}_{\rm ex}$, we have  
\begin{equation}
\rmd \op{x} =  \sum_{\varpi i, \varpi' j} \rmd x_{\varpi i, \varpi' j}
\op{c}_{\varpi i}^{\dagger} \op{c}_{\varpi' j}^{\vphantom{\dagger}} \,.
\end{equation}
The operators
\begin{equation}
\op{\beta}_n = \op{T} \op{\gamma}_n \op{T}^{-1},
\end{equation}
stemming from the transformation of the quasiparticles $\op{\gamma}_n$ of
$\ket{\Phi}$ are easily obtained from their expansion (\ref{EQ:13}) on the
set $\{\op{c}_{\varpi i}^{\dagger}\}$, the Glauber formula and the following
relations inherent to the definitions $\op{c}_{p i}^{\dagger} =
\op{c}_{i}^{\dagger}$,  $\op{c}_{h i}^{\dagger} =
\op{c}_{i}^{\vphantom{\dagger}}$
\begin{eqnarray}\label{eq:Glau}
\bigl[\op{c}_{\varpi i}^{\dagger}, \op{c}_{\varpi' j}^{\vphantom{\dagger}}\bigr]_+
&= \delta_{\varpi i, \varpi' j} \op{\unitmatrix}, \nonumber \\
\bigl[\op{c}_{\varpi i}^{\vphantom{\dagger}},
\op{c}_{\varpi' j}^{\vphantom{\dagger}}\bigr]_+ &= \sigma_{\varpi i, \varpi' j}\op{\unitmatrix},
\end{eqnarray}
where
\begin{equation}
\sigma = \left(\begin{array}{cc} 0_{d\times d}& \unitmatrix_{d\times d}\\
\unitmatrix_{d\times d} &0_{d\times d}\end{array}\right)\,\,\ \mbox{so that} \,\,  \,
\op{c}_{\varpi i}^{\vphantom{\dagger}} = \sum_{\varpi' j}
\sigma_{\varpi i, \varpi' j} \op{c}_{\varpi' j}^{\dagger}\,.
\end{equation}
Ultimately, $\op{\beta}_n$ remains linearly related to fermionic
creation/annihilation operators with an associated vector in
$\mathcal{H}^{(1)}_{\rm ex}$ given by 
\begin{equation}\label{eQ1}
\ket{\beta_n} = \exp\bigl( \rmd x - \sigma \rmd x^T \sigma\bigr) \ket{\gamma_n}.
\end{equation}
Since the operator $\op{T}$ is not necessarily unitary, the kets
$\ket{\beta_n}$  do not generally form an orthonormal family. However, they
remain orthogonal to all their partners $\ket{\bar{\beta}_n} = 
\sigma \ket{\beta_n^*}$ generated by
complex conjugation followed by the exchange of high and low 
components
\begin{equation}
\braket{\bar{\beta}_p}{\beta_n} = \bigl[\op{\beta}_p,\op{\beta}_n \bigr]_+ =
\op{T} \bigl[\op{\gamma}_p,\op{\gamma}_n \bigr]_+ \op{T}^{-1} = 0
\end{equation}
Moreover, the orthonormalization of the vectors $\ket{\beta_1},
\ldots \ket{\beta_d}$ through linear combinations 
\begin{equation}\label{EQ:ortho}
\ket{\lambda_n} = \sum_p \mathcal{O}_{np} \ket{\beta_p},
\end{equation}
automatically induces the orthonormalization of their partners 
$\ket{\bar{\lambda}_n} = \sigma \ket{\lambda^*_n}$
belonging to the orthogonal subspace. Indeed,
\begin{equation}
\braket{\bar{\lambda}_n}{\bar{\lambda}_p} = \braket{\lambda_p}{\lambda_n} =
\delta_{np}  \,\,  \, \mbox{and} \,\,  \,
\ket{\bar{\lambda}_n} = \sum_p \mathcal{O}^*_{np} \ket{\bar{\beta}_p}.
\end{equation}
As a consequence, the $2d$ vectors $\{ \ket{\lambda_n},
\ket{\bar{\lambda}_n}\}$ form an orthonormal basis of the extended one-body
space and they therefore define a new normalized HFB state $\ket{\Lambda}$. In
reality, $\ket{\Lambda}$ and $\op{T}\ket{\Phi}$ are collinear insofar as they
are a vacuum of the same quasiparticles
\begin{eqnarray}
\op{\lambda}_n = \sum_p \mathcal{O}_{np} \op{\beta}_p, \nonumber\\
\op{\lambda}_n \op{T} \ket{\Phi} = \sum_p \mathcal{O}_{np} \op{T}
\op{\gamma}_p  \ket{\Phi} = 0.
\end{eqnarray}
It should be noted that the preservation of a quasiparticle product by
application of the transformation $\op{T}$ is not at all elementary for operators
$\rmd \op{x}$ containing contributions of the type
$\op{c}^{\dagger}_i\op{c}^{\dagger}_j$. In this case, $\op{T}$ indeed changes the
particle vacuum, whereas if $\rmd \op{x}\ket{\,}=0$, $\op{T}\ket{\,}= \ket{\,}$,
and one immediately has
\begin{equation}
\op{T} \ket{\Phi} \propto \prod_{n=1}^d \op{T} \op{\gamma}_n \op{T}^{-1}
\ket{\,} = \prod_{n=1}^d \op{\beta}_n \ket{\,} = \frac{1}{\mbox{det}
  \mathcal{O}} \prod_{n=1}^d \op{\lambda}_n \ket{\,} \propto \ket{\Lambda}.
\end{equation}
In conclusion, we have shown that the vector $\exp\left(\rmd \op{x} \right)
\ket{\Phi} $ identifies, up to a multiplicative constant, to a HFB
wavefunction $\ket{\Lambda}$, that is infinitely close to $\ket{\Phi}$, and
which quasiparticles are always derived from a unitary canonical
transformation of the original fermionic operators $\op{c}_i^{\dagger}$,
$\op{c}_i^{\vphantom{\dagger}}$. $\ket{\Lambda}$ may therefore be 
noted $\ket{\Phi + \rmd \Phi}$. We are thus immediately able to determine the
elementary motion of the ansatz $\ket{W}$
\begin{eqnarray}\label{EQ:18}
\fl\ket{W+\rmd W} &=  \frac{\Pi + \rmd\Pi}{\braket{\Psi_T}{\Phi+\rmd \Phi}}\ket{\Phi+\rmd \Phi} 
\nonumber\\
\fl &= \frac{\Pi + \rmd\Pi}{\aver{\Psi_T}{\exp\left(\rmd \op{x} \right)}{\Phi}}
\exp \left( \rmd \op{x} \right)\ket{\Phi} \nonumber\\
\fl &= \biggl( 1+\frac{\rmd\Pi}{\Pi} - \langle \rmd \op{x} \rangle_{\Psi_T,\Phi} -\frac12 
\langle \rmd \op{x}^2 \rangle_{\Psi_T,\Phi} + \langle \rmd \op{x}
\rangle^2_{\Psi_T,\Phi} - \frac{\rmd\Pi}{\Pi} \langle \rmd \op{x} \rangle_{\Psi_T,\Phi} + \ldots 
\biggr) \ket{W}\nonumber \\
\fl &\quad+\biggl( 1+\frac{\rmd\Pi}{\Pi} - \langle \rmd \op{x} \rangle_{\Psi_T,\Phi} + \ldots
\biggr) \rmd \op{x} \ket{W} \nonumber\\
\fl & \quad+ \frac12 \rmd \op{x}^2 \ket{W} + \ldots
\end{eqnarray}
Here, only the contributions up to second order have been detailed and the
notation $\langle \op{A} \rangle_{\Psi_T,\Phi}$ for an operator $\op{A}$
refers to its local estimator, according to the terminology of the QMC
formalisms based on a random walk in real space
\begin{equation}\label{EQ:19}
\langle \op{A} \rangle_{\Psi_T,\Phi} = 
\frac{\aver{\Psi_T}{\op{A}}{\Phi}}{\braket{\Psi_T}{\Phi}}.
\end{equation}
At first glance and by comparison with the exact dynamics (\ref{EQ:17}), the
expansion (\ref{EQ:18}) is encouraging, showing one-body and two-body terms
applied on the ansatz $\ket{W}$. A more striking similarity can be easily
obtained by rewriting the Hamiltonian (\ref{EQ:1}) as a quadratic form of
general one-body operators (in the sense of the extended space $\mathcal{H}^{(1)}_{\rm ex}$)
\begin{equation}\label{EQ:20}
\op{H} = \op{K} - \sum_s \omega_s \op{O}^2_s,
\end{equation}
with
\begin{eqnarray}
\op{K}&= \sum_{\varpi i, \varpi' j} K_{\varpi i, \varpi' j} 
\op{c}_{\varpi i}^{\dagger} \op{c}_{\varpi' j}^{\vphantom{\dagger}},  \nonumber\\
\op{O}_s &= \sum_{\varpi i, \varpi' j} 
\left[ O_s \right]_{\varpi i, \varpi' j} 
\op{c}_{\varpi i}^{\dagger} \op{c}_{\varpi' j}^{\vphantom{\dagger}}.
\end{eqnarray}
Such an expression is immediately obtained by bringing each two-body
interaction term $\op{c}_{i}^{\dagger} \op{c}_{j}^{\dagger} 
\op{c}_{l}^{\vphantom{\dagger}} \op{c}_{k}^{\vphantom{\dagger}}$ in the one of
the forms 
$ \pm \frac12 \bigl( \op{c}_{i}^{\dagger} \op{c}_{j}^{\dagger}
  \pm \op{c}_{l}^{\vphantom{\dagger}} \op{c}_{k}^{\vphantom{\dagger}} \bigr){}^2 $,
$ \mp \frac12 \bigl( \op{c}_{i}^{\dagger} \op{c}_{l}^{\vphantom{\dagger}}
  \pm \op{c}_{j}^{\dagger} \op{c}_{k}^{\vphantom{\dagger}} \bigr){}^2 $ or
$ \pm \frac12 \bigl(\op{c}_{i}^{\dagger} \op{c}_{k}^{\vphantom{\dagger}}
  \pm \op{c}_{j}^{\dagger} \op{c}_{l}^{\vphantom{\dagger}} \bigr){}^2 $,
that may need to be completed with appropriate one-body terms. We refer to
\cite{Zha13,Lan93} for the construction, from the matrix elements $h_{i,j}$ and
$V_{i,j,k,l}$, of decompositions of the Hamiltonian that are less schematic and
more efficient for QMC treatments as they require a much lower number of
operators $\op{O}_s$. With the expression (\ref{EQ:20}) for $\op{H}$, the
exact dynamics (\ref{EQ:17}) during $\rmd\tau$ is obviously transformed into
\begin{equation}
\exp\bigl(-\rmd\tau\op{H}\bigr)\ket{W} = \ket{W} -\rmd\tau \op{K} \ket{W} +
\rmd\tau \sum_s \omega_s \op{O}^2_s \ket{W}.
\end{equation}
Thus, in the presence of two-body interactions, it can not be absorbed by a
purely determinist evolution (\ref{EQ:18}) of the ansatz $\ket{W}$: Under
this assumption, $\rmd\Pi$ and $\rmd \op{x}$ are proportional to $\rmd\tau$ and it is
impossible, to first order in $\rmd\tau$, to get back the quadratic terms of the
exact propagation. Only contributions schematically proportional to $\sqrt{\rmd\tau}$ in the 
operator $\rmd \op{x}$ are able to achieve this \textit{via} the term in $\rmd \op{x}^2$ of the 
expansion (\ref{EQ:18}). However, these contributions will also manifest themselves through one-body 
contaminations in $\sqrt{\rmd\tau}$ which have no counterpart in the evolution resulting from the 
Schr\"odinger equation. 

The idea consists in making them fluctuating with a zero
average. Mathematically, the goal is to include in $\rmd \op{x}$
(resp. $\rmd\Pi$) stochastic contributions $\rmd \op{x}_{\rm stoch}$
(resp. $\rmd\Pi_{\rm stoch}$) that depend linearly on the infinitesimal
increments $\{ \rmd W_s \} $ of independent Wiener's stochastic processes 
$\{ W_s\}$ associated with the operators $\{ \op{O}_s\}$ entering in the
decomposition (\ref{EQ:20}) of the Hamiltonian. In It\^o's calculus
\cite{Gar83}, these quantities $\rmd W_s$ indeed exhibit the properties
\begin{eqnarray}\label{EQ:22}
\mathbb{E}\left[\rmd W_s \right] = 0  \nonumber\\ 
\rmd W_s \rmd W {}_{s'}   = \delta_{s,s'} \rmd\tau, \,\, \, \forall s,s'.
\end{eqnarray}
Concretely, their simulation on a small finite time $\Delta\tau$ goes through
the introduction of random variables $\eta_s$ with zero average and
variance unity which allows to access the increments 
$\Delta W_s = \int_{\tau}^{\tau + \rmd\tau} \rmd W_s $ using $\Delta W_s = \eta_s
\sqrt{\Delta \tau}$. The variables $\eta_s$ are commonly referred to as
auxiliary fields and are often generated according to a normal
distribution. Therefore, the exact dynamics will be found by averaging the
stochastic ans\"atze $\ket{W+\rmd W}$ provided the following conditions are
simultaneously verified
\numparts\begin{eqnarray}\label{EQ:23}
\fl\frac{\rmd\Pi_{\rm det}}{\Pi} - \langle \rmd \op{x}_{\rm det} \rangle_{\Psi_T,\Phi} 
-\frac12 \langle \rmd \op{x}^2_{\rm stoch} \rangle_{\Psi_T,\Phi} +
\langle \rmd \op{x}_{\rm stoch} \rangle^2_{\Psi_T,\Phi}
- \frac{\rmd\Pi_{\rm stoch}}{\Pi}\langle \rmd \op{x}_{\rm stoch} \rangle_{\Psi_T,\Phi} 
= 0, \label{EQ:23a}\\ 
\label{EQ:23b}\rmd \op{x}_{\rm det} + \frac{\rmd\Pi_{\rm stoch}}{\Pi} \rmd \op{x}_{\rm stoch} 
- \langle \rmd \op{x}_{\rm stoch} \rangle_{\Psi_T,\Phi} \rmd \op{x}_{\rm stoch} = -\rmd\tau
\op{T}, \\ \nonumber\\
\label{EQ:23c}\frac12 \rmd \op{x}_{\rm stoch}^2 =\rmd\tau \sum_s \omega_s \op{O}^2_s,
\end{eqnarray}\endnumparts
where $\rmd \op{x}_{\rm det}$ (resp. $d{\Pi}_{\rm det}$) stands for the
determinist terms, proportional to $\rmd\tau$, which form $\rmd \op{x}$ (resp. $d{\Pi}$). These 
relations do not uniquely determine the equations of motion for the quasiparticle states and for 
the prefactor $\Pi$. In view of the properties (\ref{EQ:22}), the last constraint (\ref{EQ:23c}) is 
satisfied as long as
\begin{equation}\label{EQ:25}
\rmd \op{x}_{{\rm stoch}} = \sum_s \sqrt{2\omega_s} \op{O}_s \rmd W_s \,.
\end{equation}
Moreover, by making explicit the linear relation between $\rmd\Pi_{\rm stoch}$
and the increments $\rmd W_s$ in the form 
$\rmd\Pi_{{\rm stoch}} = \Pi \sum_s \sqrt{2\omega_s} g_s \rmd W_s $ with arbitrary
scalars $g_s$, we get successively from the equalities 
(\ref{EQ:23a} - \ref{EQ:23b})
\begin{eqnarray}\label{eQ2}
\rmd \op{x}_{\rm det} &= - \rmd\tau \left[ \op{T} -2 \omega_s 
\langle \op{O}_s -g_s \rangle_{\Psi_T,\Phi} \op{O}_s \right] \nonumber\\ 
\rmd\Pi_{\rm det}  &= - \Pi \, \rmd\tau \, \langle \op{H} \rangle_{\Psi_T,\Phi} \,.
\end{eqnarray}
Eventually, by iterating the process for each ansatz $\ket{W+\rmd W}$ obtained at
the end of the propagation during $\rm\rmd\tau$, we are led to a representation
(\ref{EQ:psig}) of the ground state. The quasiparticles of the HFB vacua
$\ket{\Phi_{\tau}}$ are defined by the realizations at the imaginary time $\tau
\rightarrow \infty$ of a Brownian motion in the extended one-body space, which is directly deduced 
from equations (\ref{eQ1}, \ref{EQ:25}, \ref{eQ2}, and \ref{EQ:ortho})
\begin{eqnarray}\label{EQ:27}
\ket{\gamma_n + d\gamma_n} = \mathcal{O}\biggl\{ \ket{\gamma_n} &- \rmd\tau \Bigl[
\tilde{K} - \sum_s \omega_s \Bigl( \tilde{O}^2_s + 
2 \langle \op{O}_s -g_s \rangle_{\Psi_T,\Phi}  \tilde{O}_s \Bigl) \Bigr]
\ket{\gamma_n} \biggr. \nonumber\\
 \biggl.&+ \sum_s \sqrt{2\omega_s} \rmd W_s \tilde{O}_s \ket{\gamma_n}
\biggr\}.
\end{eqnarray}
Here, $\mathcal{O}$ is a formal notation to indicate an orthonormalization
process according to the previous discussion and $\tilde{K} = K - \sigma K^T
\sigma$, $\tilde{O}_s = O_s - \sigma O^T_s \sigma$. The prefactor $\Pi$
evolves according to
\begin{equation}\label{EQ:28}
\frac{\rmd\Pi}{\Pi} = - \rmd\tau \langle \op{H} \rangle_{\Psi_T,\Phi} + \sum_s
\sqrt{2\omega_s} g_s \rmd W_s .
\end{equation}
This stochastic differential equation easily fits into It\^o's formalism and its
solution $\Pi_{\tau}$ at time $\tau$ is given by
\begin{equation}\label{EQ:29}
\fl\Pi_{\tau} = \braket{\Psi_T}{\Phi_0} \exp \biggl\{-\!\!\int_0^{\tau} \!\Bigl[
\rmd\tau' \Bigl(\langle \op{H} \rangle_{\Psi_T,\Phi_{\tau'}} + \sum_s
\omega_s g_{s,\tau'}^2 \Bigr)
+ \sum_s\sqrt{2\omega_s} g_{s,\tau'} \rmd W_{s,\tau'}\Bigr]\biggr\} .
\end{equation}
The quantities $g_s$ thus remain undefined on average by the
conditions (\ref{EQ:23a}, \ref{EQ:23b}, \ref{EQ:23c}), which sometimes leads
to qualify them as 
``stochastic gauges'' \cite{gauges}. They are however involved in the growth of
the averaged quadratic error and thus can affect the efficiency of the
sampling. Up to date, no numerical applications involving Slater determinants
have taken into account an imaginary-time dependence of the gauges $\{g_s\}$,
and the choices $g_s = 0$ or $g_s = \langle \op{O}_s \rangle_{\Psi_T,\Psi_T} $
have been proposed \cite{eeam}.

It should be finally noted that the Brownian motion (\ref{EQ:27}, \ref{EQ:28}) of HFB walkers 
allows to find the usual reconstruction scheme (\ref{EQ:psig}) of the ground state with stochastic 
Slater determinants. Indeed, let us consider such a Hartree-Fock (HF) wavefunction 
$\ket{\Phi} = \prod_{n=1}^{N} \op{c}^{\dagger}_{\phi_n}\ket{\,} $
built from $N$ orthonormal one-body states $\{ \ket{\phi_n}, n= 1, 2, \ldots, N
\}$. By completing their set with a family of unoccupied vectors 
$\{ \ket{\bar{\phi}_{\nu}}, \nu = 1, \ldots, d-N\}$
to have an orthonormal basis of $\mathcal{H}^{(1)}$, $\ket{\Phi}$ appears as a
vacuum for the operators $\op{c}^{\dagger}_{\phi_n}$ and
$\op{c}^{\vphantom{\dagger}}_{\bar{\phi}_{\nu}} $ which play the role of
quasiparticles $\op{\gamma}_1, \ldots, \op{\gamma}_d$. The associated vectors
in the extended one-body space $\mathcal{H}^{(1)}_{\rm ex}$ are therefore
given by
\begin{equation}
\fl\ket{\gamma_n} = \left( \begin{array}{c}
\ket{\phi_n}\\ 0 \end{array}\right) \,\, \mbox{for }\, n= 1, \ldots, N, \,\,\
\mbox{and }\,\,
\ket{\gamma_n} = \left(\begin{array}{c}
0\\ \ket{\bar{\phi}^*_{n-N}} 
\end{array}\right)\, \,\mbox{for }\, n= N+1, \ldots, d .
\end{equation}
By solely retaining in the rewriting (\ref{EQ:20}) of the Hamiltonian
operators $\op{O}_s$ that conserve the particle number, this structure is
preserved by the dynamics (\ref{EQ:27}), which immediately gives the
evolution of the occupied states $\ket{\phi_n}$. The results obtained this way
are identical to those presented in the original reference
\cite{Zha03} in the limit of a continuous imaginary time. Indeed, the
strategy adopted in this work is based on a discretization of the exact
propagation which is then reformulated in terms of a random walk \textit{via} the
Hubbard-Stratonovich transformation. The method that we have followed here
actually originates from the QMC approaches developed for systems of
interacting bosons by stochastic extension of the mean-field approximations
\cite{Car01}. 

\section{Control of the phase and infinite-variance problems} \label{sec:variance}

Apart from a few exceptional models, the QMC reconstruction of the
ground state $\ket{\Psi_G}$ suffers from the pathological phase problem (or
sign problem when the walkers can be restricted to maintain a wavefunction with
real components during their Brownian motion). In practice, this problem
manifests itself by an exponential growth, with the time and/or the size of
the system, of the statistical error on the averaged value of any
observable. Its origin is intimately connected to the principles of quantum
physics that set the state vector $\ket{\Psi_G}$ up to a phase. The
probability distribution of walkers being real and positive, it may not be
sensitive to this phase and inexorably samples the vectors
$\{\rme^{\rmi\Theta_G} \ket{\Psi_G}, \Theta_G \in \mathbb{R}\}$ which are physically
equivalent. From then on, one immediately sees that by averaging all the
realizations, regardless of the phase $\Theta_G $ to which they lead for the
ground state, some mutually cancel each other and alter the efficiency of
the reconstruction scheme. The missing access to $\Theta_G $ prevents an exact
control of this problem, except from a modification of the
stochastic dynamics of walkers to ensure an overlap of constant phase with
$\ket{\Psi_G}$. 

Let us concretely consider the previously developed QMC scheme
(\ref{EQ:psig}), that is guided by a trial state $\ket{\Psi_T}$. Its viability
is actually questioned as soon as even the smallest population of realizations 
$\ket{W_{\tau^*}} = \Pi_{\tau^*}\ket{\Phi_{\tau^*}}/\braket{\Psi_T}{\Phi_{\tau^*}}$,  
which 
collectively have a zero average overlap with $\ket{\Psi_G}$, emerges at any
given time $\tau^*$. Their statistical weight remains however negligible at
this time and no problem may actually be detected. Noting
$\ket{\Psi_{\tau^*}^{\perp}}$ the sum of these specific walkers, we have
$\braket{\Psi_G}{\Psi_{\tau^*}^{\perp}} = 0$, so that at any later time $\tau$
\begin{equation}
\fl\aver{\Psi_G}{\exp\bigl[-(\tau-\tau^*) \op{H} \bigr]}{\Psi_{\tau^*}^{\perp}} =
\mathbb{E}\left[\braket{\Psi_G}{W_{\tau}}\right] =
\exp\left[-(\tau-\tau^*) E_G \right] \braket{\Psi_G}{\Psi_{\tau^*}^{\perp}} 
\end{equation}
($E_G$ is the energy of the exact ground state). Therefore, the set of
paths stemming from the problematic realizations at a time $\tau^*$ forms a
population characterized by a zero-averaged overlap with $\ket{\Psi_G}$. Thus,
the number of walkers not contributing to the sampling increases with the
imaginary time and their presence only deteriorates the signal-to-noise
ratio. At the same time, the proportion of realizations truly
participating to the reconstruction of the ground state decreases
exponentially. With the dynamics (\ref{EQ:27}, \ref{EQ:28}), such
problems occur inexorably when the phase of the multiplicative prefactor $\Pi$
changes during the Brownian motion. From then on, the overlaps
$\braket{\Psi_G}{W_{\tau^*}}$ are distributed in the entire complex plane and
the formation of a pathological sum $\ket{\Psi_{\tau^*}^{\perp}}$ can not be
avoided. Indeed,
\begin{equation}
\fl\braket{\Psi_G}{W_{\tau^*}} \propto \lim_{\tau \to \infty} 
\aver{\Psi_T}{\exp\bigl[-(\tau-\tau^*) \op{H} \bigr]}{W_{\tau^*}} = 
\lim_{\tau \to \infty} \mathbb{E}\left[\braket{\Psi_T}{W_{\tau}}\right]= 
\lim_{\tau \to \infty} \mathbb{E}\left[\Pi_{\tau}\right] .
\end{equation}
Here, we assumed $\braket{\Psi_T}{\Psi_G} \neq 0$, so that the exact
ground state can be seen as the propagation of the trial wavefunction
$\ket{\Psi_T}$ 
after a very long time. In view of the expression (\ref{EQ:29}) of the
factor $\Pi$, a phase problem arises with the sampling (\ref{EQ:psig}, \ref{EQ:27}, \ref{EQ:28}) as 
soon as some coefficients $\omega_s <0$ are
required. This situation is in fact systematic for all realistic Hamiltonian
rewritten as a quadratic form of one-body operators \cite{Zha13, Lan93}. It
follows that we necessarily have complex components for the quasiparticle
states and therefore a local energy $\langle \op{H} \rangle_{\Psi_T,\Phi}$
with a complex value inducing a variable phase for $\Pi_{\tau}$ (even in the
absence of gauges). However, it should be noted that in the utopian assumption
where walkers would be guided by the exact ground state $\ket{\Psi_T} =
\ket{\Psi_G}$, the sampling would not display any phase problem: The factors
$\Pi_{\tau}$ would have a constant phase since they would all be given by
$\exp\left(-\tau E_G \right) \braket{\Psi_G}{\Phi_0}$, with the choice
$g_s=0$. With an incorrect state $\ket{\Psi_T}$, this observation naturally
leads to a control of the phase problem \textit{via} the use of \textit{biased}
multiplicative factors $\tilde{\Pi}_{\tau}$ coming from a dynamics
(\ref{EQ:28}) where the local energy is replaced by its real part (and
similarly for the terms related to the stochastic gauges). The stochastic
reformulation is thus no longer equivalent to the Schr\"odinger equation in
imaginary time and only an approximation $\ket{\tilde{\Psi}_G}$ to the ground state
can be reached
\begin{equation}
\ket{\tilde{\Psi}_G} \propto \lim_{\tau \to \infty} 
\mathbb{E}\left[\tilde{\Pi}_{\tau} 
\frac{\ket{\Phi_{\tau}}}{\braket{\Psi_T}{\Phi_{\tau}} }\right],
\end{equation}
with
\begin{equation}\label{EQ:33}
\fl\tilde{\Pi}_{\tau} = \braket{\Psi_T}{\Phi_0}\exp\biggl\{-\!\int_0^{\tau}\!
\Bigl[ \rmd\tau' \Bigl( \Re \bigl( \langle \op{H} \rangle_{\Psi_T,\Phi_{\tau'}}\bigr)
+ \frac12 \sum_s R^2_{s,\tau'} \Bigr) + \sum_s R_{s,\tau'} \rmd W_{s,\tau'} \Bigr]\biggr\},
\end{equation}
where we set $R_{s} = \Re{\left(\sqrt{2\omega_s} g_{s}\right)} $. Note that by
adjusting the phase of the trial wavefunction, the factors $\tilde{\Pi}_{\tau}$
are easily made real and positive. Thus, they identify to the weights of the
vectors $\ket{\tilde{W}_{\tau}} = \ket{\Phi_{\tau}}/
\braket{\Psi_T}{\Phi_{\tau}}$ generated by the Brownian motion
(\ref{EQ:27}) and they can be simply absorbed by replacing the sampled
probability distribution $\mathbb{P}(\ket{\Phi}, \tilde{\Pi}, \tau)$ with
$\tilde{\mathbb{P}} \propto \tilde{\Pi} \mathbb{P}$. In the following, we
will note $\mathbb{E}_{\tilde{\Pi}}[\ldots]$ the averages evaluated with this modified distribution.

\begin{figure}[t!]
\begin{center}
\includegraphics[width=\linewidth]{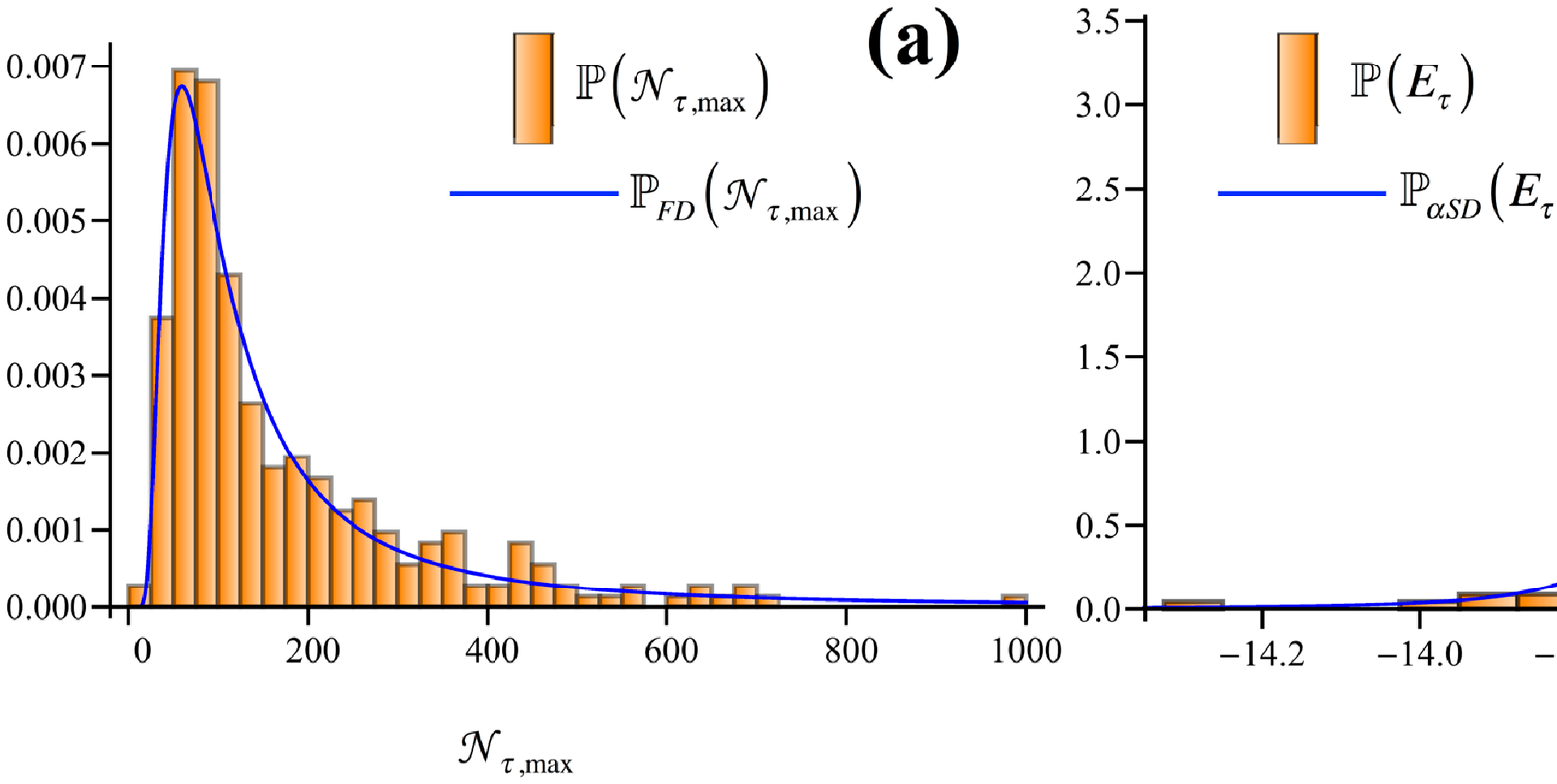}
\end{center}
\caption{(Color online) Emergence of infinite-variance problems in the QMC
scheme (\ref{EQ:27}, \ref{EQ:28}) when only real stochastic
wavefunctions are generated. We consider the Hubbard model (\ref{EQ:114})
on a  $4 \times 4$ half-filled cluster in the repulsive regime $U=4t$ with HF
walkers. The single-particle states remain real during their Brownian motion
as long as the interaction term on a site ${\bf r}$  is cast into the form
$2\op{n}_{{\bf r}\uparrow} \op{n}_{{\bf r}\downarrow}  = \op{n}_{{\bf r}\uparrow} +
\op{n}_{{\bf r}\downarrow} - \left( \op{n}_{{\bf r}\uparrow} -
\op{n}_{{\bf r}\downarrow}\right)^2$.
300 independent populations, with the number of walkers fixed to $N_w = 100$,
are obtained through a
semi-implicit Euler algorithm and an adaptive step control technique to solve
the stochastic differential equations (\ref{EQ:27}). The sampling
according to the weight $\Pi_{\tau}$  is performed with a reconfiguration
scheme detailed in \cite{Sor98}. For each population, the averaged energy
$E_{\tau} = \mathbb{E}_{\Pi}\bigl[\langle \op{H} \rangle_{\Psi_T,\Phi_{\tau}}\bigr]$
and the norm $\mathcal{N}_{\tau} = 
\bigl|\bigl|\ket{\tilde{W}_{\tau}}\bigr|\bigr| = 1/|\braket{\Psi_T}{\Phi_{\tau}}|$
of generated walkers 
$\ket{\tilde{W}_{\tau}} = \ket{\Phi_{\tau}}/\braket{\Psi_T}{\Phi_{\tau}}$  
are calculated as a function of the
imaginary time $\tau$. According to the extreme-value theorem \cite{Kot00}, a
power-law tail $\mathcal{N}_{\tau}^{-1-\nu}$ for the distribution of
$\mathcal{N}_{\tau}$ maps into
a Fr\'echet law $\mathbb{P}_{FD}(\mathcal{N}_{\tau, {\rm max}}) \propto
\rme^{-(\mathcal{N}_{\tau, {\rm max}}/S)^{-\nu}} $ for the maximum value 
$\mathcal{N}_{\tau, {\rm max}}$ over a finite and sufficiently large
sequence of realizations ($S$ is a scale parameter).
(a) Histogram of the empirical probability distribution function 
$\mathbb{P}(\mathcal{N}_{\tau, {\rm max}})$ of $\mathcal{N}_{\tau, {\rm max}}$
at imaginary time $\tau=5/t$. The fit to the expected distribution 
$\mathbb{P}_{FD}(\mathcal{N}_{\tau, {\rm max}})$ is shown by the continuous
line. The associated $p$-value of 0.675 in a Pearson $\chi^2$ test confirms
the validity of the Fr\'echet law hypothesis. An exponent $\nu \simeq 3.94 >
2$ is thus extracted and indicates an infinite variance (\ref{EQ:34})
of the error on the exact many-body state.
(b)~Histogram of the empirical probability distribution function
$\mathbb{P}(E_{\tau})$ of the energy at imaginary time $\tau=5/t$ . The results
are not normally distributed ($p$-value $\simeq 0.04$), but rather follows the
asymmetric L\'evy stable distribution $\mathbb{P}_{\alpha SD}(E_{\tau})$ of index
$\alpha \simeq 1.776$ ($p$-value $\simeq 0.576$).  Such behavior agrees with
the generalized central-limit theorem for the sum of independent random
variables of infinite variance. Consequently, the standard Monte-Carlo
estimate $\bar{E}_{\tau} = -13.483(8)t$, for the energy $E_{\tau}$ averaged over all populations, is meaningless. Without the knowledge of an infinite-variance
problem, a systematic error compared to the exact ground-state energy $E_G =
-13.62t$ is highlighted.} 
\label{fig:0}
\end{figure}

Even if the resulting stochastic scheme is, by construction, free of the phase
problem, its applicability in Monte-Carlo simulations still requires a finite
dispersion of the realizations $\ket{\tilde{W}_{\tau}}$  around their average
$\ket{\tilde{\Psi}_{\tau}} = \mathbb{E}_{\tilde{\Pi}}[\ket{\tilde{W}_{\tau}}]$ for
all imaginary time $\tau$
\begin{equation}\label{EQ:34}
\fl\mbox{Var}_{\tilde{\Pi}}\left[ \ket{\tilde{W}_{\tau}} \right] = 
\mathbb{E}_{\tilde{\Pi}}\left[ \left| \left| \ket{\tilde{\Psi}_{\tau}} -
\ket{\tilde{W}_{\tau}}  \right| \right|^2 \right] = 
\mathbb{E}_{\tilde{\Pi}}\left[ 
\frac{1}{\left| \braket{\Psi_T}{\Phi_{\tau}}\right|^2 } \right] - \left|
\left| \ket{\tilde{\Psi}_{\tau}} \right| \right|^2 < \infty .
\end{equation}
This condition guarantees the convergence towards the approximated
ground state $\ket{\tilde{\Psi}_G}$ in the limit of an infinite number of
paths 
$\mathcal{N}_r$, the statistical error decreasing with $\mathcal{N}_r$ like
$1/\sqrt{\mathcal{N}_r}$. Unfortunately, a divergence of the variance (\ref{EQ:34}) is clearly 
expected as soon as a significant number of
walkers $\ket{\Phi_{\tau}}$ are generated in directions almost orthogonal to
the trial state $\ket{\Psi_T}$. The elimination of such pathological
realizations in order to obtain an applicable QMC formalism requires the use
of an additional approximation supplementing the one controlling the phase
problem (\ref{EQ:33}). Zhang \& Krakauer precisely proposed to take
advantage of the phase of the overlap  $\braket{\Psi_T}{\Phi_{\tau}}$ for each
elementary step of the Brownian motion to detect a potential proximity of the
walker to the hypersurface $\braket{\Psi_T}{\Phi_{\tau}} = 0 $ \cite{Zha03}. As
part of the exact stochastic reformulation (\ref{EQ:27}, \ref{EQ:28}) and
thanks to It\^o's rules, it is easy to check that the infinitesimal evolution
$\ket{\Phi} \rightarrow \ket{\Phi + \rmd \Phi}$ of a HFB wavefunction $\ket{\Phi}$
induces a variation of its overlap with the trial state according to
\begin{eqnarray}\label{EQ:35}
\frac{\braket{\Psi_T}{\Phi+\rmd \Phi}}{\braket{\Psi_T}{\Phi}} = 
\frac{\braket{\Phi}{\Phi+\rmd \Phi}}{\aver{\Phi}{\exp\left(\rmd \op{x}\right)}{\Phi}}
\frac{\rmd\Pi}{\Pi} \, &\exp \Bigl( \rmd\tau \sum_s \omega_s \langle \op{O}_s - 
g_s \rangle_{\Psi_T,\Phi}^2 \nonumber\\
&\Bigl.+ \sum_s \sqrt{2 \omega_s} \langle \op{O}_s 
- g_s \rangle_{\Psi_T,\Phi} \rmd W_s \Bigr) \,,
\end{eqnarray}
where the first term comes precisely from the relation $\ket{\Phi+\rmd \Phi}
\propto \exp\left(\rmd \op{x}\right)\ket{\Phi}$. For $ \braket{\Psi_T}{\Phi}
\rightarrow 0$, the contributions using the local estimators $\langle \op{O}_s
\rangle_{\Psi_T,\Phi}$ dominate by diverging as $1/ \braket{\Psi_T}{\Phi}$. Among
these, one can also only retain the fluctuating contributions insofar as they
vary schematically as $\sqrt{\rmd\tau}$ whereas the others are proportional to
$\rmd\tau$. Finally, after noting that the introduction of the biased factors
$\tilde{\Pi}$ in place of $\Pi$ is equivalent to neglecting
$\Im{\left(\rmd\Pi/\Pi\right)}$, the phase $\rmd\theta$ of the ratio (\ref{EQ:35}) is then 
approximately given by
\begin{equation}\label{EQ:36}
\rmd\theta \sim \sum_s \Im\Bigl(\sqrt{2 \omega_s} \langle \op{O}_s - g_s
    \rangle_{\Psi_T,\Phi} \Bigr) \rmd W_s \sim \mathcal{O} 
\left(\frac{\sqrt{\rmd\tau}}{\braket{\Psi_T}{\Phi}}
\right) \,.
\end{equation}
As a consequence, a sudden phase shift $\rmd\theta$ can be reasonably expected
when the quasiparticle vacuum $\ket{\Phi}$ is close to orthogonality with
$\ket{\Psi_T}$. The emergence of a sampling of infinite variance can therefore
be avoided by changing the dynamics of the biased weight $\tilde{\Pi}$  of a
walker: The more its overlap's phase with the trial state varies during
$\rmd\tau$, the more its biased weight is reduced. Concretely, we will follow
Zhang \& Krakauer's strategy by requesting the biased weights $\tilde{\Pi}$
to evolve as
\begin{equation}\label{EQ:37}
\fl\frac{\tilde{\Pi}_{\tau +\rmd\tau}}{\tilde{\Pi}_{\tau}} =
\max\biggl\{ 0, \cos \frac{\braket{\Psi_T}{\Phi_{\tau 
+\rmd\tau}}}{\braket{\Psi_T}{\Phi_{\tau}}}\biggr\}\!
\exp \biggl[-\rmd\tau \Re\Bigl( \langle \op{H} \rangle_{\Psi_T,\Phi_{\tau}} + \!\frac12
    \sum_s \!R_{s,\tau}^2\Bigr) + \sum_s \!R_{s,\tau} W_{s,\tau}\biggr] .
\end{equation}
This choice leads to exclude from the sampling the realizations undergoing, in
the complex plane associated to $\braket{\Psi_T}{\Phi}$, a phase shift
$\rmd\theta \geq \pi/2$ during their motion in a time $\rmd\tau$. In this plane
and at the limit $\tau \to \infty$, corresponding to the reconstruction of the
ground state, it results that the region around the real axis and far from the
origin is almost exclusively populated. Note finally that the relation (\ref{EQ:37}) must be 
understood in the sense of a numerical
implementation, $\rmd\tau$ being simply replaced by a small finite time step
$\Delta\tau$. Mathematically, we point out that, to first order in
$\rmd\theta$, $\cos{\rmd\theta}$ and $\exp\left(-\rmd\theta^2/2 \right)$ are equal
in view of the expression (\ref{EQ:36}) of the phase shift $\rmd\theta$. As a
consequence, the differential equation satisfied by the biased weight
$\tilde{\Pi}$ to constrain the phase and infinite-variance problems should rather be written as
\begin{equation}\label{EQ:38}
\fl\frac{d\tilde{\Pi}}{\tilde{\Pi}} = -\rmd\tau \Re{\left( \langle \op{H} 
\rangle_{\Psi_T,\Phi_{\tau}} \right)} - \frac{\rmd\tau}{2} \sum_s 
\left[ \Im{\left( \sqrt{2 \omega_s} \langle \op{O}_s - g_s
  \rangle_{\Psi_T,\Phi}\right)} \right]^2 + \sum_s R_{s} \rmd W_{s} \,.
\end{equation}
For a Hamiltonian (\ref{EQ:1}) defined from real elements of a one-body
matrix $h_{i,j}$ and a two-body matrix $V_{i,j,k,l}$, the operators $\op{K}$ and
$\op{O}_s$, which make it possible to write it in the quadratic form (\ref{EQ:20}), can always be 
chosen so that they admit a real representation in the basis $\{\ket{\varpi i}\}$ of 
$\mathcal{H}^{(1)}_{\rm ex}$. In very rare cases, the associated 
coefficients $\omega_s$ are all positive. The Hubbard model, which allows one
to highlight the generic properties of a fermionic system on a lattice, is an
example that meets such conditions both in the attractive and repulsive
regimes, irrespective of the dimensionality of the lattice. The QMC
reformulation with guided dynamics (\ref{EQ:psig}, \ref{EQ:27}, \ref{EQ:28})
then encounters no phase problem provided it is initiated with a HFB state
which matrices $U$ and $V$ are real. Moreover, one should also be able to
write the trial state 
$\ket{\Psi_T}$ as linear combination of such HFB vacua with real
amplitudes. With these conditions, the Brownian motion preserves the real
character of the Bogoliubov transformation at all imaginary time and the
multiplicative factors $\Pi$ have a constant phase. It is nevertheless not
guaranteed to obtain the exact ground state as an infinite-variance
problem may arise   
when realizations $\ket{\Phi}$ with $ \braket{\Psi_T}{\Phi}
\rightarrow 0$ are generated. For them, equation (\ref{EQ:35}) leads to a ratio
dominated by $\exp \bigl( \sum_s \sqrt{2 \omega_s} \langle \op{O}_s -
g_s \rangle_{\Psi_T,\Phi} \rmd W_s \bigr)$, which diverges or vanishes
depending on the sign of $\langle \op{O}_s - g_s \rangle_{\Psi_T,\Phi}
\rmd W_s$. Therefore, the infinitesimal motion may take the walker away, or on
the contrary, bring it closer to the origin of the complex plane of the
overlap with the trial state. In addition, no approximated QMC scheme
can be recovered through the use of the biased weights (\ref{EQ:37})
because $ \braket{\Psi_T}{\Phi}$ strictly remains on the real axis throughout
the evolution. Hence, as soon as the dynamics only explores real Bogoliubov
transformations, no reliability can be granted to the approach
(\ref{EQ:psig}, \ref{EQ:27}, \ref{EQ:28}). With Slater determinants to guide
and start the Brownian motion, we actually showed \cite{Bon12} that the QMC
scheme considered here is equivalent to the samplings proposed in 2004
\cite{Cor04} and 2007 \cite{Jui07} for the Hubbard model and free from sign
problems. For small cells, numerical simulations have quickly highlighted
systematic errors \cite{Cor08} that we have linked to the emergence of an
infinite variance for the exact state \cite{Bon12}. An illustration is given
in figure~\ref{fig:0}. 

To conclude, it should be noted that infinite-variance problem is not specific
to the dynamics (\ref{EQ:27}) considered here. The standard auxiliary-field
approach (\ref{EQ:16}), which 
explores Slater determinants and differs by the absence
of local estimators $\langle \op{O}_s \rangle_{\Psi_T,\Phi}$ in the
determinist evolution, generally suffers from a similar pathology when
importance sampling is included and is exempt of sign problems \cite{Shi16}.
\section{Estimates of observables}\label{sec:estobs}
\subsection{General considerations}\label{sec:gencon}

As an illustrative example, consider the determination of the energy $E_G$ of
the ground state $\ket{\Psi_G}$. As part of an exact propagation (\ref{EQ:15}) in  imaginary
time, we immediately obtain
\begin{equation}
E_G = \lim_{\tau \to \infty} 
\frac{\Re{\aver{\Psi_T}{\op{H}\exp\bigl(-\tau \op{H} \bigr)}{\Phi_0}}}
{\Re{\aver{\Psi_T}{\exp\bigl(-\tau \op{H} \bigr)}{\Phi_0}}},
\end{equation}
provided the initial state $\ket{\Phi_0}$ and the trial state
$\ket{\Psi_T}$ are not orthogonal to $\ket{\Psi_G}$. With the stochastic
interpretation (\ref{EQ:psig}) of the dynamics where the walkers are
generated according to the importance of their overlap with $\ket{\Psi_T}$, 
\begin{equation}
E_G = \lim_{\tau \to \infty} 
\frac{\mathbb{E}\left[\Re{\left(\Pi_{\tau} \langle \op{H} \rangle_{\Psi_T,\Phi_{\tau}}\right)}
\right]}{\mathbb{E}\Bigl[\Re{\left(\Pi_{\tau}\right)}\Bigr]}
\end{equation}
follows. In practice, only one approached state $\ket{\tilde{\Psi}_G}$ is however
accessible through the introduction of real positive biased weights
$\tilde{\Pi}$ in place of the multiplicative factors $\Pi$, and the associated
energy $\tilde{E}_G$ will be evaluated according to 
\begin{equation}\label{EQ:41}
\tilde{E}_G = \lim_{\tau \to \infty} 
\frac{\mathbb{E}\left[\tilde{\Pi}_{\tau}\Re{\left( \langle \op{H} 
\rangle_{\Psi_T,\Phi_{\tau}}\right)}\right]}
{\mathbb{E}\left[\tilde{\Pi}_{\tau}\right]} = 
\lim_{\tau \to \infty} \mathbb{E}_{\tilde{\Pi}}\left[\Re{\left(
      \langle \op{H} \rangle_{\Psi_T,\Phi_{\tau}}\right)}\right] \,.
\end{equation}
The elimination at all times of the walkers close to orthogonality with the trial state guarantees 
a finite variance for this energy and thus ensures the validity of its reconstruction by the 
Monte-Carlo techniques. Indeed,
\begin{equation}
\fl\mbox{Var}_{\tilde{\Pi}} \Bigl[\Re{\Bigl(\langle \op{H} 
\rangle_{\Psi_T,\Phi_{\tau}}\Bigr)}\Bigr]\! =\! \mathbb{E}_{\tilde{\Pi}}\left[\!\left\{\tilde{E}_G 
-  \Re{\left( \langle \op{H} \rangle_{\Psi_T,\Phi_{\tau}}\right)}\right\}^2\!\right]\! = \!
\mathbb{E}_{\tilde{\Pi}}\left[\left\{ \Re{\left( \langle \op{H} 
\rangle_{\Psi_T,\Phi_{\tau}}\right)}\right\}^2\right] -
\tilde{E}_G^2,
\end{equation}
with $\bigl( \Re{\langle \op{H} \rangle_{\Psi_T,\Phi_{\tau}}}\bigr)^2 \leq
\bigl| \langle \op{H} \rangle_{\Psi_T,\Phi_{\tau}} \bigr|^2 \leq
\aver{\Psi_T}{\op{H}^2}{\Psi_T} \braket{\tilde{W}_{\tau}}{\tilde{W}_{\tau}}$
as a consequence of the Schwartz inequality, and therefore
\begin{equation}\label{EQ:infty}
\mbox{Var}_{\tilde{\Pi}} \left[\Re{\left(
      \langle \op{H} \rangle_{\Psi_T,\Phi_{\tau}}\right)}\right] 
\leq \aver{\Psi_T}{\op{H}^2}{\Psi_T}\mbox{Var}_{\tilde{\Pi}} \bigl[
  \ket{\tilde{W}_{\tau}}\bigr] - \tilde{E}_G^2 < \infty \,.
\end{equation}
These considerations extend to any observable $\op{A}$ commuting with the
Hamiltonian. In the opposite case, the analogous estimator for the energy
(\ref{EQ:41}) (obtained by replacing $\op{H}$ by $\op{A}$) is commonly
called mixed estimator $\tilde{A}_G^{({\rm mix})}$ of the observable. It can
only offer an approximation to the true averaged value $\tilde{A}_G =
\aver{\tilde{\Psi}_G}{\op{A}}{\tilde{\Psi}_G}$ in the biased ground state
through the following relation, valid only to first order in the
difference
\begin{equation}
\tilde{A}_G \approx 2 \tilde{A}_G^{({\rm mix})} -\aver{\Psi_T}{\op{A}}{\Psi_T} \,.
\end{equation}
Direct simulation of $\tilde{A}_G$ is however possible through the ``back-propagation''
technique introduced in references \cite{Zha97, Pur04}. Its principle is
based again on the results established in section~\ref{sec:sto}, which we will
write here in terms of the infinitesimal ``left'' propagation of a 
dyad $\op{\sigma} = \Pi \ket{\Phi} \bra{\Psi_T}/\braket{\Psi_T}{\Phi} $
built from a HFB vacuum and the trial state
\begin{equation}
\exp\bigl(-\rmd\tau \op{H} \bigr)\op{\sigma} = \mathbb{E}\left[\left( \Pi + \rmd\Pi \right) 
\frac{\exp\left(\rmd \op{x}\right) \ket{\Phi} \bra{\Psi_T}}
{\aver{\Psi_T}{\exp\left(\rmd \op{x}\right)}{\Phi}} \right]\,,
\end{equation}
where $\rmd\Pi$, $\rmd \op{x}$ satisfy the conditions (\ref{EQ:23a}-\ref{EQ:23c}). One 
immediately verifies that an elementary modification of the bra $\bra{\Psi_T}$ with the same 
operators $\exp\left(\rmd \op{x}\right)$ also leads to a ``right'' propagation of
$\op{\sigma}$: 
\begin{equation}
\mathbb{E}\left[ \left( \Pi + \rmd\Pi \right) 
\frac{\ket{\Phi} \bra{\Psi_T}\exp\left(\rmd \op{x}\right) }
{\aver{\Psi_T}{\exp\left(\rmd \op{x}\right)}{\Phi}} \right] =
\op{\sigma} \exp\bigl(-\rmd\tau \op{H} \bigr) \,.
\end{equation}
The projection of the trial wavefunction $\ket{\Psi_T}$ on the ground state 
can therefore be achieved by reusing, however in reverse order, the stochastic
transformations $\exp\left(\rmd \op{x}\right)$ successively undergone by the
initial HFB vector $\ket{\Phi_0}$ during its Brownian motion guided by
$\ket{\Psi_T}$. This evolution in imaginary time of $\ket{\Psi_T}$ is precisely required to 
access the expectation values in the ground state. Indeed,
\begin{equation}
A_G = \aver{{\Psi}_G}{\op{A}}{{\Psi}_G} = \lim_{\tau_K,\tau_B \to \infty}
\frac{\Re{\aver{\Psi_T}{\exp\bigl(-\tau_B \op{H} \bigr) \op{A}\,\exp\bigl(-\tau_K 
\op{H}\bigr)}{\Phi_0}}}
{\Re{\aver{\Psi_T}{\exp\bigl(-\tau_B \op{H} \bigr)\exp\bigl(-\tau_K \op{H} \bigr)}{\Phi_0}}} 
\,.
\end{equation}
The QMC reconstruction of this ``true estimator'' is thus simply based on a
prolongation of the motion of the HFB walkers to the time $\tau_K +
\tau_B$. However, the practical implementation of the method also needs to be
capable to apply the exponential of the general one-body operators $\rmd \op{x}$
on the considered trial state. Using an approximation $\ket{\Phi'_{0}}$ of HFB type
for $\ket{\Psi_T}$ is therefore natural. With such a choice
to guide the realizations and constrain them through biased weights, we are
finally able to completely characterize the approached ground state
$\ket{\tilde{\Psi}_G}$ by determining the expectation values of interest
$\tilde{A}_G$ according to
\begin{equation}\label{EQ:48}
\tilde{A}_G = \lim_{\tau_K,\tau_B \to \infty}
\frac{\mathbb{E}\Bigl[ \tilde{\Pi}_{\tau_K + \tau_B} \Re{\langle \op{A}
  \rangle_{\Phi'_{\tau_B},\Phi_{\tau_K}}}\Bigr]}
{\mathbb{E}\Bigl[\tilde{\Pi}_{\tau_K + \tau_B} \Bigr]}\,.
\end{equation}
Here, $\tilde{\Pi}$ and the quasiparticles $\{\op{\gamma}_n\}$ of the walkers
$\ket{\Phi}$ must be determined up to time $\tau_K + \tau_B$ \textit{via} the
evolution equations (\ref{EQ:27}, \ref{EQ:37}) with the trial state
$\ket{\Psi_T}=\ket{\Phi'_{0}}$. The HFB vacuum $\ket{\Phi_{\tau_K}}$ is
defined from its intermediate solution at time $\tau_K$ whereas the HFB
wavefunction $\ket{\Phi'_{\tau_B}}$ results from a random walk during
$\tau_B$, starting from $\ket{\Phi'_{0}}$. Its quasiparticles are precisely
obtained from the following Langevin equation in the extended one-body space
\begin{eqnarray}\label{EQ:49}
\fl\ket{\gamma'_n}_{\tau + \rmd\tau} = \mathcal{O}&\biggl\{\ket{\gamma'_n}_{\tau} &-
\rmd\tau \Bigl[ \tilde{K}^{\dagger} -  \sum_s \omega_s \Bigl(\tilde{O}_s^{\dagger 2}
+ 2 \langle \op{O}_s - g_s \rangle^*_{\Phi'_{0},\Phi_{\tau_K + \tau_B - \tau}} 
\tilde{O}_s^{\dagger} \Bigr) \Bigr] \ket{\gamma'_n}_{\tau} \biggr. \nonumber\\
& \biggl. &+\sum_s \left( \sqrt{2 \omega_s}\right)^* \rmd W_{s,\tau_K + \tau_B - \tau} 
\tilde{O}_s^{\dagger}\ket{\gamma'_n}_{\tau} \biggr\},
\end{eqnarray}
where the local estimators
$\langle \op{O}_s  \rangle_{\Phi'_{0},\Phi_{\tau_K + \tau_B - \tau}} $ and the
increments of Wiener's processes $ \rmd W_{s,\tau_K + \tau_B - \tau}$ are those
used during the complementary motion of the ket $\ket{\Phi}$ between times
$\tau_K$ and $\tau_K + \tau_B$.

In the end, the applicability of the extension of the ``phaseless QMC''
formalism to HFB states (\ref{EQ:27}, \ref{EQ:37}, \ref{EQ:41}, \ref{EQ:48}, \ref{EQ:49}) is 
essentially based on the knowledge of the
local estimators $\langle \op{A}  \rangle_{\Phi',\Phi} $ of observables between
two Bogoliubov vacua $\ket{\Phi}$, $\ket{\Phi'}$. The overlaps
$\braket{\Phi'}{\Phi}$ are also needed to bias the
weights of the realizations so as to control the variance. We proceed
to the determination of these quantities in the following paragraphs.

\subsection{Extended Wick's theorem}
\label{sec:ewt}
Wick's theorem plays a key role in theoretical treatments of interacting
fermionic systems using Slater determinants or HFB wavefunctions. It allows
one to express the expectation value of any operator in a vacuum of particles
or quasiparticles in terms of the normal and abnormal elementary contractions
\cite{Wic50, Bla85}. We show in this section that the result actually remains
valid for local estimators of operators between two non-orthogonal HF or HFB
factorized states. We also prove that the functional of binary contractions
corresponds to the expansion of a Pfaffian, which greatly facilitates the
numerical evaluation for operators with more than two bodies. 

With two Slater determinants, this extension of Wick's theorem to
matrix elements is at the heart of auxiliary fields QMC
approaches as well as ``phaseless QMC'' simulations. Its well known proof is
based on the same principle as the one usually presented for the expectation
values \cite{Low55, Bla85}. The hybrid case
of a matrix element between a HF state and a HFB wavefunction has
never been considered in a general way to our knowledge, even if some
partial results have been reported \cite{Car11}. Yet, it is required, for
example, in the previous QMC scheme applied to HF walkers guided with
a trial HFB wavefunction to absorb at least approximately the pairing
correlations. Here we propose a general demonstration of Wick's
theorem for the local estimators, valid regardless of the HF or HFB
nature of each of the two vacua. It is inspired by the work of
Balian \& Br\'ezin about non-unitary Bogoliubov transformations
\cite{Bal69}, as well as Gaudin's work about Wick's theorem at finite
temperature for a fermionic system without interaction \cite{Gau60}.

Let $\ket{\Phi}$ and $\ket{\Phi'}$ be two non-orthogonal states, each being a
Slater determinant or a Bogoliubov vacuum. Our aim is to
determine the matrix element between these two wavefunctions, of a product of
operators which factors are \textit{arbitrary} linear combinations of
fermionic creation $\op{c}^{\dagger}_{{i}}$ and annihilation
$\op{c}^{\phantom{\dagger}}_{{i}}$ operators. In other words, it is
to express:
\begin{equation}\label{eq:54}
\aver{\Phi'}{\op{Q}_1\ldots \op{Q}_M}{\Phi} = \braket{\Phi'}{\Phi} \mbox{Tr}
\left[ \op{Q}_1\ldots \op{Q}_M \ket{\Phi} \bra{\tilde{\Phi}'} \right],
\end{equation}
with
\begin{equation}
\bra{\tilde{\Phi}'} = \bra{\Phi'} / \braket{\Phi'}{\Phi}, \,\,\,\mbox{and}
\end{equation}
\begin{equation} 
\op{Q}_a = \sum_{{i}} 
\left(\op{c}^{\dagger}_{{i}} Y^*_{{i},a}
+ \op{c}^{\phantom{\dagger}}_{{i}} X^*_{{i},a} \right),
\,\,\,\forall a=1, \ldots,M.
\end{equation}
Let us denote $\{ \op{\gamma}_n^{\phantom{\dagger}} \}$ and 
$\{ \op{\gamma}_n'^{\phantom{\dagger}} \}$ the HF or HFB quasiparticle
operators (with $1\leq n \leq d$) associated to their respective vacua
$\ket{\Phi}$ and $\ket{\Phi'}$. The corresponding orthonormal bases of the
one-body 
extended space are designated by $\{ \ket{\gamma_n},  \ket{\bar{\gamma}_n}\}$
and $\{ \ket{\gamma'_n},  \ket{\bar{\gamma}_n'}\}$. Using the expansions
(\ref{EQ:13}, \ref{EQ:14}), the $d\times d$ matrices defined by the overlaps
$\mathbb{F}_{mn} = \braket{\gamma'_m}{\gamma_n}$ and $\mathbb{G}_{mn} =
\braket{\gamma'_m}{\bar{\gamma}_n}$ allow to relate the two families of
quasiparticle through
\begin{eqnarray} \label{eq:55}
\op{\gamma}'_m &= \sum_{\varpi {i}} \op{c}_{\varpi {i}}^{\dagger} 
\braket{\varpi {i}}{\gamma'_m} = \sum_{\varpi {i}}
\sum_{n=1}^{d} \left( \op{\gamma}_n \braket{\gamma_n}{\varpi {i}} +
\op{\gamma}_n^{\dagger} \braket{\bar{\gamma}_n}{\varpi {i}}
\right)\braket{\varpi {i}}{\gamma'_m} \nonumber\\
&= \sum_{n=1}^{d} \left(\mathbb{F}_{mn}^* \op{\gamma}_n + \mathbb{G}_{mn}^* \op{\gamma}_n^{\dagger} 
\right).
\end{eqnarray}
Let us note that equation (\ref{eq:55}) can also be obtained directly by expanding
$\ket{\gamma'_m}$  in the basis $\{ \ket{\gamma_n},  \ket{\bar{\gamma}_n}\}$
and using the linearity of the operators $\op{\gamma}'^{\phantom{\dagger}}_m$, 
$\op{\gamma}_n^{\phantom{\dagger}}$,
and $\op{\gamma}_n^{\dagger}$  in terms of the states $\ket{\gamma'_m}$, 
$ \ket{\gamma_n}$, and $ \ket{\bar{\gamma}_n}$, respectively. Thus, the
matrices $\mathbb{F}^{T}$ and  $\mathbb{G}^{T}$  are analogous to the matrices
$U$ and $V$ of the Bogoliubov transformation (\ref{eq:9p}) that define
$\ket{\Phi'}$ when the vacuum of the original fermionic operators 
$\{ \op{c}_{{i}}^{\phantom{\dagger}} \}$ is replaced by the vacuum
$\ket{\Phi}$, the one of the quasiparticles 
$\{ \op{\gamma}_n^{\phantom{\dagger}} \}$.

We show now that the matrix $\mathbb{F}$ is necessarily invertible as a result
of the nonorthogonality of the states $\ket{\Phi}$ and $\ket{\Phi'}$. In the
case of two Slater determinants, $\mathbb{F}$ is reduced to a block-diagonal
matrix:  One of them, $f$, of size $N\times N$, contains the scalar products
between the occupied one-body states in both wavefunctions, and the
other one, $\bar{f}$, of size $(d-N)\times (d-N)$, entails the overlaps
between empty one-body states. Besides, $\braket{\Phi'}{\Phi}$ is easily
obtained from the  anticommutation relation
$\bigl[\op{c}^{\phantom{\dagger}}_{\phi'_m}, \op{c}_{\phi_n}^{\dagger}
\bigr]_+ = \braket{\phi'_m}{\phi_n}\, \op{\unitmatrix}$, for two arbitrary
individual occupied states $\ket{\phi'_m}$ and $\ket{\phi_n}$  respectively
related to the determinants $\ket{\Phi'}$ and
$\ket{\Phi}$: $\braket{\Phi'}{\Phi} = \mbox{det} f$. However,
each of the two HF states can alternatively be obtained, up to a phase factor,
starting from the fully filled one-body space: It suffices to annihilate the
fermions occupying the one-body states $\ket{\bar{\phi}'_{\mu}}$  for
$\ket{\Phi'}$, and $\ket{\bar{\phi}_{\nu}}$  for $\ket{\Phi}$, which should
actually be empty in these determinants. Still using the anticommutator 
$\bigl[\op{c}_{\bar{\phi}'_{\mu}}^{\dagger}, 
\op{c}^{\phantom{\dagger}}_{\bar{\phi}_{\nu}}
\bigr]_+ = \braket{\bar{\phi}'_{\mu}}{\bar{\phi}_{\nu}} \op{\unitmatrix}$,
one now obtains $\braket{\Phi'}{\Phi} = \rme^{\rmi \theta} \mbox{det} \bar{f}^*$
where the global phase $\rme^{\rmi \theta}$ is determined from the components
$\left\{ \phi'_{{i},n} \right\}$ and 
$\left\{ \phi_{{i},n} \right\}$. One ends with
\begin{equation}\label{eq:56}
|\braket{\Phi'}{\Phi}|^2 = |\mbox{det} \mathbb{F}| .
\end{equation}

Equation~(\ref{eq:56}) remains actually valid with two Bogoliubov vacua
\cite{Oni66} or in the hybrid case of a HF wavefunction and  a HFB
state. Without loss of generality and to simplify the discussion, we will
temporarily assume $\ket{\Phi'}$  to be a coherent pair state. We can then
repeat the reasoning of section~\ref{sec:HFB} to write the state
$\ket{\Phi'}$ in a 
BCS-type form, namely $\ket{\Phi'} = \rme^{\rmi \theta} \prod_{\alpha=1}^{d/2}
\bigl(f_{\alpha} + g_{\alpha} \op{\Gamma}_{\alpha}^{\dagger}
\op{\Gamma}_{\tilde{\alpha}}^{\dagger} \bigr) \ket{\Phi} $, using the
Bloch-Messiah-Zumino decomposition (\ref{eq:15-15p}) of the matrices
$\mathbb{F}^T$ and $\mathbb{G}^T$, coming from the Bogoliubov transformation
(\ref{eq:55}). Here, the operators $\op{\Gamma}_{\alpha}^{\dagger}$ and 
$\op{\Gamma}_{\tilde{\alpha}}^{\dagger}$ are linear combinations of the
quasiparticles $\op{\gamma}_n^{\dagger}$ (which play the role of creation
operators $\op{c}_{{i}}^{\dagger}$ in the usual case of a
Bogoliubov transformation applied to a fermion vacuum); $\rme^{\rmi \theta}$ is again a
phase factor and the positive real numbers $f_{\alpha}$, $g_{\alpha}$ define the
canonical form of matrices $\mathbb{F}^T$, $\mathbb{G}^T$. Hence,
$|\braket{\Phi'}{\Phi}| = \prod_{\alpha=1}^{d/2} f_{\alpha} $, and Onishi's
equation (\ref{eq:56}) is recovered, since $|\mbox{det} \mathbb{F}| =
\prod_{\alpha=1}^{d/2} f_{\alpha}^2$ due to the Bloch-Messiah-Zumino
factorization of the matrix  $\mathbb{F}^T$. Finally, the non orthogonality of 
$\ket{\Phi}$ and $\ket{\Phi'}$ guarantees the matrix $\mathbb{F}$ to be
invertible, irrespective of whether they are of  HF or HFB type.

Let us now  introduce a non-unitary transformation of the Fock space 
\begin{equation}
\op{\mathcal{T}} = \exp\left( \sum_{m=1}^{d} \sum_{n=m+1}^{d} 
\left(\mathbb{F}^{-1} \mathbb{G}\right)_{mn} \op{\gamma}_m \op{\gamma}_n \right)\,.
\end{equation}
We first show that the matrix $\mathbb{F}^{-1} \mathbb{G}$ is
antisymmetric. This follows both from the fermionic algebra of the sets of
quasiparticles 
$\left\{ \op{\gamma}_n \right\}$ and $\left\{ \op{\gamma}'_n \right\}$ as well
as from the linearity of equation (\ref{eq:55}). Using
$\left[\op{\gamma}'^{\dagger}_{m}, \op{\gamma}'^{\dagger}_{n}\right]_+ = 
\left(\mathbb{F} \mathbb{G}^T + \mathbb{G} \mathbb{F}^T \right)_{mn}
\op{\unitmatrix} = 0$ yields $\mathbb{F}^{-1} \mathbb{G} = 
-\mathbb{G}^T \left(\mathbb{F}^T \right)^{-1} = 
-\left(\mathbb{F}^{-1} \mathbb{G} \right){}^T $. 
In addition, $\op{\mathcal{T}}$ leaves invariant the
annihilation operators of the quasiparticle in the state  $\ket{\Phi}$, and the
creation operators are transformed into linear combinations of their
counterparts entering in the second wavefunction $\ket{\Phi'}$. We will note 
$\gopp{\gamma}{m}$ these combinations
\begin{eqnarray}\label{eq:58}
\op{\mathcal{T}}^{-1} \op{\gamma}_m \op{\mathcal{T}}  &= \op{\gamma}_m ,\nonumber\\
 \op{\mathcal{T}}^{-1}\op{\gamma}_m^{\dagger} \op{\mathcal{T}}   &= 
\gopp{\gamma}{m}= \op{\gamma}_m^{\dagger} -\frac12
\sum_{n,n'=1}^{d}\left(\mathbb{F}^{-1} \mathbb{G} \right)_{nn'}
\left[\op{\gamma}_n\op{\gamma}_{n'},\op{\gamma}_m^{\dagger}\right] \nonumber\\
& =\op{\gamma}_m^{\dagger} + \sum_{n=1}^{d}\left(\mathbb{F}^{-1}
  \mathbb{G} \right)_{mn}\op{\gamma}_n = \sum_{n=1}^{d}\left(\mathbb{F}^{-1}
   \right)_{mn} \op{\gamma}_n'^{\dagger}\,.
\end{eqnarray}
It is important to note that the so defined operators meet by construction the
canonical anticommutation relations although they are not connected by the hermitian conjugation:
\begin{equation}\label{eq:59}
\bigl[\op{\gamma}_m,\gopp{\gamma}{n}\bigr]_+ =
\op{\mathcal{T}}^{-1}\left[\op{\gamma}_m ,\op{\gamma}_n^{\dagger}\right]_+
\op{\mathcal{T}} = \delta_{m,n} \op{\unitmatrix}\,.
\end{equation}

The transformation $\op{\mathcal{T}}$ also induces a non-trivial  resolution of the
identity in the Fock space. Consider the orthonormal basis formed by
the states $\ket{n_1 \ldots n_d}$ of either the HF or the HFB vacuum
$\ket{\Phi}$ by creating one or more of its quasiparticle
\begin{equation}\label{eq:60}
\ket{n_1 \ldots n_d} = \left(\op{\gamma}_1^{\dagger}\right)^{n_1}  \ldots
\left(\op{\gamma}_d^{\dagger}\right)^{n_d} \ket{\Phi} ,
\end{equation}
with $n_l = 0$ or $1$ ($l=1, \ldots, d$). Applying the inverse transformation
$\op{\mathcal{T}}^{-1}$ on these vectors, the obtained kets $\rket{n_1 \ldots n_d}
$ are no longer orthonormal and now correspond to the creation of quasiparticles
$\gopp{\gamma}{m}$ on the state $\ket{\Phi}$:
\begin{eqnarray}\label{eq:61}
\rket{n_1 \ldots n_d} &= \op{\mathcal{T}}^{-1} \ket{n_1 \ldots n_d} = \left( \op{\mathcal{T}}^{-1} 
\op{\gamma}_1^{\dagger} \op{\mathcal{T}}\right)^{n_1}  \ldots 
\left( \op{\mathcal{T}}^{-1} \op{\gamma}_d^{\dagger} 
\op{\mathcal{T}}\right)^{n_d} \ket{\Phi} \nonumber\\
&= \left(\gopp{\gamma}{1} \right)^{n_1}   \ldots 
\left(\gopp{\gamma}{d} \right)^{n_d} \ket{\Phi}\,.
\end{eqnarray}
Here we used the invariance under $\op{\mathcal{T}}$ or
$\op{\mathcal{T}}^{-1}$ of the 
vacuum $\ket{\Phi}$ for the operators $\op{\gamma}_n$. The states (\ref{eq:61}) are also right 
eigenvectors of the non-hermitian operators $\gopp{\gamma}{l} \op{\gamma}_l$ resulting from the 
transformation of the quasiparticle numbers $\op{n}_l^{\phantom{\dagger}} = \op{\gamma}_l^{\dagger} 
\op{\gamma}_l^{\phantom{\dagger}}$ associated to $\ket{\Phi}$,
\begin{equation}
\gopp{\gamma}{l} \op{\gamma}_l \rket{n_1 \ldots n_d} =
\op{\mathcal{T}}^{-1} \op{\gamma}_l^{\dagger} \op{\mathcal{T}}
\op{\mathcal{T}}^{-1} \op{\gamma}_l^{\phantom{\dagger}} \op{\mathcal{T}}
\op{\mathcal{T}}^{-1} \ket{n_1 \ldots n_d} = n_l \rket{n_1 \ldots n_d}\,.
\end{equation}
The adjoint basis, formed by the left eigenvectors $\lbra{n_1 \ldots n_d}$, is
immediately found according to the same developments, except that 
the dual vectors of the occupation number representation (\ref{eq:60}) are
transformed under $\op{\mathcal{T}}$
\begin{eqnarray}
\lbra{n_1 &\ldots& n_d} = \bra{n_1 \ldots n_d} \op{\mathcal{T}} = \bra{\Phi}
\op{\mathcal{T}} \left( \op{\mathcal{T}}^{-1} \op{\gamma}_d^{\phantom{\dagger}} 
\op{\mathcal{T}}\right)^{n_d} \ldots \left( \op{\mathcal{T}}^{-1} 
\op{\gamma}_1^{\phantom{\dagger}} \op{\mathcal{T}}\right)^{n_1} \nonumber\\
\label{eq:62t}&\quad&\,\,\,\,\,\,\,\, =
\bra{\tilde{\Phi}'} \left(\op{\gamma}_d^{\phantom{\dagger}}\right)^{n_d}
\ldots \left(\op{\gamma}_1^{\phantom{\dagger}}\right)^{n_1} ,\\
\lbra{n_1 &\ldots& n_d}  \gopp{\gamma}{l}
\op{\gamma}_l^{\phantom{\dagger}} = \bra{n_1 \ldots n_d} \op{\mathcal{T}}
\op{\mathcal{T}}^{-1} \op{n}_l^{\phantom{\dagger}} 
\op{\mathcal{T}} = \lbra{n_1 \ldots n_d} n_l\,.
\label{eq:63}
\end{eqnarray}
These are therefore quasiparticles $\op{\gamma}_m^{\dagger}$ which are created
on the second vacuum $\ket{\tilde{\Phi}'}$. 
To achieve such a result, $\op{\mathcal{T}}$ should however connect the two
considered coherent HF or HFB states : $\bra{\Phi} \op{\mathcal{T}} =
\bra{\tilde{\Phi}'}$. 
The proof simply consists in noticing that 
$\bra{\tilde{\Phi}'} \op{\mathcal{T}}^{-1} \op{n}_l = 
\bra{\tilde{\Phi}'} \gopp{\gamma}{l}
\op{\gamma}_l^{\phantom{\dagger}}
\op{\mathcal{T}}^{-1} = 0$ since $\gopp{\gamma}{l}$ (see equation (\ref{eq:58}))
is indeed a linear combination of the operators $\op{\gamma}'^{\dagger}_{m}$,
with $\bra{\tilde{\Phi}'} = \bra{\Phi'} / \braket{\Phi'}{\Phi}$ being their
associated vacuum. As a result, the vectors 
$\bra{\tilde{\Phi}'} \op{\mathcal{T}}^{-1}$ and 
$\bra{\Phi}$ are necessarily collinear, as they both correspond to the
configuration where all occupation numbers $\{n_l\}$ are zero. Since
$\aver{\tilde{\Phi}'}{\op{\mathcal{T}}^{-1}}{\Phi} =
\braket{\tilde{\Phi}'}{\Phi} = 1$, one thus obtains 
$\bra{\tilde{\Phi}'} \op{\mathcal{T}}^{-1} = \bra{\Phi}$. This result
identifies to Thouless' theorem in 
its most general form \cite{Tho60,Rin80}. It therefore remains valid irrespective of the HF or
HFB nature of the two non-orthogonal involved vacua. Finally, we can
summarize the previous results through the closure relation and the
bi-orthogonality relation satisfied by the vectors (\ref{eq:61}) and (\ref{eq:62t}) stemming from 
the Thouless transformation of the occupation-number representation
\begin{eqnarray}\label{eq:65}
\fl\sum_{\{ n_1 \ldots n_d \} } {\rket{n_1 \ldots n_d}} {\lbra{n_1 \ldots n_d}} =
\op{\mathcal{T}}^{-1} \sum_{ \{ n_1 \ldots n_d \} } \ket{n_1 \ldots n_d} 
\bra{n_1 \ldots n_d} \op{\mathcal{T}} = \op{\unitmatrix},\\
\fl\lrbraket{n_1 \ldots n_d}{n_1' \ldots n_d'}= \aver{n_1 \ldots
  n_d}{\op{\mathcal{T}}\op{\mathcal{T}}^{-1}}{n_1' \ldots n_d'} =
\delta_{n_1,n_1'} \ldots \delta_{n_d,n_d'}\,.
\end{eqnarray}

Therefore, the dyad $\ket{\Phi} \bra{\tilde{\Phi}'}$, necessary to estimate
the matrix elements (\ref{eq:54}), can be easily extracted by eliminating
all configurations with at least one quasiparticle excitation. Thanks to the
previous algebraic developments, such a goal is achieved \textit{via} a Gaussian
non-hermitian operator 
\begin{equation}\label{eq:66}
\op{\mathcal{D}} = \frac{1}{\mathcal{Z}}\exp{\left( -\sum_{l=1}^d \vartheta_l 
\gopp{\gamma}{l} \op{\gamma}_l^{\phantom{\dagger}}\right)},
\end{equation}
in which the real numbers $\{\vartheta_l \}$ are arbitrary and $\mathcal{Z}$
ensures normalization. Physically, $\op{\mathcal{D}}$ simply results from the
Thouless transformation of the density operator describing (in the
grand-canonical ensemble) the equilibrium state of an ideal quasiparticle gas 
$\{\op{\gamma}_l \}$. In this interpretation, the parameters 
$\vartheta_l$ are therefore linked to the individual energies $\epsilon_l$ of
these quasiparticles, to the temperature $\beta^{-1}$, and to the chemical
potential $\mu$  
according to $\vartheta_l = \beta(\epsilon_l-\mu)$. Besides, the Gaussian
operator $\op{\mathcal{D}}$ is diagonal in the representation (\ref{eq:65}) of the right and left 
eigenvectors of the operators $\{\gopp{\gamma}{l} \op{\gamma}_l^{\phantom{\dagger}} \}$
\begin{equation}
\op{\mathcal{D}} = \sum_{ \{ n_1 \ldots n_d \} } \rket{n_1 \ldots n_d}
\frac{\rme^{-\vartheta_1 n_1}}{1+\rme^{-\vartheta_1}} \ldots 
\frac{\rme^{-\vartheta_d n_d}}{1+\rme^{-\vartheta_d}} \lbra{n_1 \ldots n_d}\,.
\end{equation}
In the limit $\vartheta_1 \to \infty, \ldots, \vartheta_d \to
\infty$ (corresponding to zero temperature and a chemical
potential $\mu<\min_l(\epsilon_l)$  for the underlying  perfect gas), only the
configuration $n_1=0, \ldots , n_d=0$ remains and  $\op{\mathcal{D}}$ therefore
reduces to the dyad $\rket{0 \ldots 0} \lbra{0 \ldots 0}= \ket{\Phi} \bra{\tilde{\Phi}'}$.  
Eventually, we can thus bring any matrix element between two non-orthogonal states, each of HF or
HFB type, to an expectation value in the Gaussian ansatz (\ref{eq:66})
\begin{equation}
\aver{\Phi '}{\op{Q}_1 \ldots \op{Q}_M}{\Phi} =
\braket{\Phi '}{\Phi} \lim_{\{ \vartheta_l \to \infty \} } 
\langle \op{Q}_1 \ldots \op{Q}_M \rangle_{\op{\mathcal{D}}}\,,
\end{equation}
where $\langle \ldots \rangle_{\op{\mathcal{D}}} = 
\mbox{Tr} (\ldots \op{\mathcal{D}})$ denotes the expectation value
in a  statistical mixture represented by the density operator
$\op{\mathcal{D}}$. 

Let us now first focus on the simple cases where either $\op{Q}_1 =
\gopp{\gamma}{l}$ or $\op{Q}_1 = \op{\gamma}_{l}$. One can show that
\begin{equation}
\gopp{\gamma}{l} \op{\mathcal{D}} = \rme^{\vartheta_l} \op{\mathcal{D}}
\gopp{\gamma}{l},\quad \op{\gamma}_{l} \op{\mathcal{D}} = 
\rme^{-\vartheta_l} \op{\mathcal{D}} \op{\gamma}_{l},
\end{equation}
by integrating
\begin{equation}\label{eq:70p}
\frac{\partial \bigl(\op{\mathcal{D}}^{-1} \op{Q}_1 \op{\mathcal{D}} \bigr)}
{\partial \vartheta_l} = \op{\mathcal{D}}^{-1} 
\bigl[\gopp{\gamma}{l}\op{\gamma}_{l},\op{Q}_1 \bigr] \op{\mathcal{D}} = 
\pm \bigl(\op{\mathcal{D}}^{-1} \op{Q}_1 \op{\mathcal{D}} \bigr),
\end{equation}
that directly follows from the Gaussian form of $\op{\mathcal{D}}$ as well as
from the anticommutation relations of the operators $\gopp{\gamma}{l}$ and
$\op{\gamma}_{l}$. In equation (\ref{eq:70p}), the signs $(+)$ and  $(-)$ refer to 
$\op{Q}_1 = \gopp{\gamma}{l}$ and $\op{Q}_1 = \op{\gamma}_{l}$,
respectively. Under these circumstances, the cyclic invariance of the trace
allows to relate the two averaged values 
$\langle \op{Q}_1 \op{Q}_2 \ldots \op{Q}_M\rangle_{\op{\mathcal{D}}}$ and 
$\langle \op{Q}_2 \ldots \op{Q}_M \op{Q}_1  \rangle_{\op{\mathcal{D}}}$ as
\begin{eqnarray}
\langle \op{Q}_2 \ldots \op{Q}_M \op{Q}_1\rangle_{\op{\mathcal{D}}}&=
\mbox{Tr}\left(\op{Q}_2 \ldots \op{Q}_M \op{Q}_1 \op{\mathcal{D}} \right)=\rme^{\pm \vartheta_l} 
\mbox{Tr}\left(\op{Q}_2 \ldots \op{Q}_M \op{\mathcal{D}} \op{Q}_1  \right) \nonumber \\
&= \rme^{\pm \vartheta_l} \mbox{Tr}\left(\op{Q}_1 \op{Q}_2 \ldots \op{Q}_M\op{\mathcal{D}} \right)
= \rme^{\pm \vartheta_l} \langle \op{Q}_1 \op{Q}_2 \ldots \op{Q}_M\rangle_{\op{\mathcal{D}}}.
\end{eqnarray}
Assuming from now on that $M$ is even, the calculation of the expectation
value of their product with the Gaussian density operator, possibly
non-hermitian, is therefore determined through a recursive procedure defined
by 
\begin{equation}\label{eq:72}
\langle \op{Q}_1 \op{Q}_2 \ldots \op{Q}_M\rangle_{\op{\mathcal{D}}} =
\sum_{b=2}^{M} (-1)^b \frac{\bigl[\op{Q}_1,\op{Q}_b
  \bigr]_+}{1+\rme^{\pm \vartheta_l}} \langle \op{Q}_2 \ldots \op{Q}_{b-1}
\op{Q}_{b+1}\ldots \op{Q}_M\rangle_{\op{\mathcal{D}}}  .
\end{equation}
When the number of factors is as small as two, we obtain the following
contractions $\langle \op{Q}_1 \op{Q}_2 \rangle_{\op{\mathcal{D}}}$:  
\begin{equation}\label{eq:73}
\langle \op{Q}_1 \op{Q}_2 \rangle_{\op{\mathcal{D}}} = 
\frac{\bigl[\op{Q}_1,\op{Q}_2 \bigr]_+}{1+\rme^{\pm \vartheta_l}}.
\end{equation}
Let us recall that, at this stage, the operator $\op{Q}_1$ is limited to
either $ \gopp{\gamma}{l}$ (+ sign) or $\op{\gamma}_{l}$ (- sign). On the other
hand, $\op{Q}_2$ can be any linear combination of fermionic
elementary operators. The expansion (\ref{eq:72}) can thus obviously be
written in a linear form in the operators $\op{Q}_1$ considered up to now
\begin{equation}\label{eq:74}
\langle \op{Q}_1 \op{Q}_2 \ldots \op{Q}_M\rangle_{\op{\mathcal{D}}} =
\sum_{b=2}^{M} (-1)^b \langle \op{Q}_1 \op{Q}_b \rangle_{\op{\mathcal{D}}}
\langle \op{Q}_2 \ldots \op{Q}_{b-1}
\op{Q}_{b+1}\ldots \op{Q}_M\rangle_{\op{\mathcal{D}}}\,.
\end{equation}
The generalization to any factor is immediate as long as it can be linearly
expanded in terms of the quasiparticles $\{\op{\gamma}_{l}\}$ and $
\{\gopp{\gamma}{l}\}$.  To achieve this, the kets $\{ \ket{\gamma_{l}} \}$ and 
$\bigl\{ \ket{\bar{\tilde{\gamma}}'_{l}} = \sum_{m=1}^d
\left(\mathbb{F}^{-1}\right)_{lm} \ket{\bar{\gamma}'_{m}} \bigr\} $, on which
these operators depend linearly (see  equation (\ref{eq:58})), must form a basis. They
define two non-orthogonal subspaces and their total number is equal to $2d$, the
dimension of $\mathcal{H}_{\rm ex}^{(1)}$. Hence, it just needs to be checked
that they are linearly independent. If $\sum_{m=1}^{d} 
\left( \lambda_m \ket{\gamma_m} + \lambda'_m
\ket{\bar{\tilde{\gamma}}'_m} \right) = 0$
where $\lambda_m$ and $\lambda'_m$ are scalar numbers, the following equations
necessarily hold true
\begin{eqnarray}
\sum_{m=1}^{d} \left( \lambda_m \braket{\gamma'_l}{\gamma_m}+ \lambda'_m
  \braket{\gamma'_l}{\bar{\tilde{\gamma}}'_m} \right) &=
\sum_{m=1}^{d}\mathbb{F}_{lm} \lambda_m = 0, \nonumber\\
\sum_{m=1}^{d} \left( \lambda_m \braket{\bar{\gamma}_l}{\gamma_m}+ \lambda'_m
  \braket{\bar{\gamma}_l}{\bar{\tilde{\gamma}}'_m} \right) &=
\sum_{m,n=1}^{d} \lambda'_m \left(\mathbb{F}^{-1}\right)_{mn}
\braket{\bar{\gamma}_l}{\bar{\gamma}'_n} = 0\,.
\end{eqnarray}
The matrix $\mathbb{F}$ being invertible, the coefficients $\lambda_m$ are
therefore zero. 
Returning to the amplitudes $\left\{U_{{i}, l}, V_{{i}, l}\right\}$  and 
$\left\{U'_{{i}, m}, V'_{{i}, m}\right\}$
of the canonical Bogoliubov transformations defining  the two considered coherent
states $\ket{\Phi}$ and $\ket{\Phi'}$, respectively, it turns out that the
overlaps $\braket{\bar{\gamma}^{\phantom{*}}_l}{\bar{\gamma}'_n}$
are given by the matrix transposed of $\mathbb{F}$
\begin{equation}
\braket{\bar{\gamma}^{\phantom{*}}_l}{\bar{\gamma}'_n} = 
\braket{U^{\phantom{*}}_l}{U'_n} +  \braket{V^{\phantom{*}}_l}{V'_n} = 
\braket{U'^*_n}{U^*_l} + \braket{V'^*_n}{V^*_l} =
\braket{\gamma'_n}{\gamma^{\phantom{*}}_l} = \mathbb{F}^{\phantom{*}}_{nl}\,.
\end{equation}
As a result, $ \sum_{m,n=1}^{d} \lambda'_m \left(\mathbb{F}^{-1}\right)_{mn}
\braket{\bar{\gamma}_l}{\bar{\gamma}'_n} = \lambda'_l = 0 $ and the set $\left\{ \ket{\gamma_{l}},
\ket{\bar{\tilde{\gamma}}'_{l}}  \right\} $ is indeed complete. 
These developments show additionally that the 
adjoint basis consists of the bras $\bigl\{ \bra{\tilde{\gamma}'_{l}} =
\sum_{m=1}^d \left(\mathbb{F}^{-1}\right)_{lm} \bra{\gamma'_{m}}, 
\bra{\bar{\gamma}_{l}} \bigr\}$. We may thus
finally write the completeness relation in the extended one-body space
\begin{equation}\label{eq:77}
\sum_{l=1}^{d} \bigl( \ket{\gamma^{\phantom{*}}_l} \bra{\tilde{\gamma}'_l} +  
\ket{\bar{\tilde{\gamma}}'_l} \bra{\bar{\gamma}_l} \bigr) = \unitmatrix .
\end{equation}
By attaching the two usual vectors 
\begin{equation}
\ket{Q_a} = \left(\begin{array}{c} \ket{Y^*_a}\\ \ket{X^*_a}
\end{array}\right) \,\,\, \mbox{and}\,\,\,  \ket{\bar{Q}_a} = \left(\begin{array}{c}
  \ket{X_a}\\ \ket{Y_a} \end{array} \right),
\end{equation}
to each operator $\op{Q}_a = \sum_{{i}} \bigl( 
\op{c}^{\dagger}_{{i}} Y^*_{{i}, a} +
\op{c}^{\phantom{\dagger}}_{{i}} X^*_{{i}, a} \bigr) $, 
this resolution of the identity, together with (\ref{EQ:13}), induces the following expansions from the
decomposition of the ket $\ket{Q_a}$ or the bra $\bra{\bar{Q}_a}$
\begin{equation}\label{eq:78}
\op{Q}^{\phantom{*}}_a = \sum_{l=1}^{d}\left( \op{\gamma}^{\phantom{*}}_l 
\braket{\tilde{\gamma}'_l}{Q^{\phantom{*}}_a} 
+ \gopp{\gamma}{l} \braket{\bar{\gamma}^{\phantom{*}}_l}{{Q}^{\phantom{*}}_a} \right) = 
\sum_{l=1}^{d}\left(
\braket{\bar{Q}^{\phantom{*}}_a}{\gamma^{\phantom{*}}_l} \gopp{\gamma}{l} +
\braket{\bar{Q}^{\phantom{*}}_a}{\bar{\tilde{\gamma}}_l} \op{\gamma}^{\phantom{*}}_l
\right) .
\end{equation}
Therefore, any first factor $\op{Q}_1$ can always be reduced to a linear
combination of the quasiparticle operators $\gopp{\gamma}{l}$  and 
$\op{\gamma}_{l}$. The relation (\ref{eq:74}), giving a recursive expression of the 
expectation value $\langle \op{Q}_1 \op{Q}_2 \ldots \op{Q}_M
\rangle_{\op{\mathcal{D}}}$ in a Gaussian density operator, is therefore valid
in general, and corresponds to Wick's theorem. It only requires the knowledge
of binary contractions $\langle \op{Q}_a \op{Q}_b 
\rangle_{\op{\mathcal{D}}}$ that are, moreover, obtained by combining the
expansions (\ref{eq:78}), the previously obtained (\ref{eq:73}) elementary contractions
$\langle \gopp{\gamma}{l} \op{Q}_b \rangle_{\op{\mathcal{D}}}$, $\langle \op{\gamma}_l \op{Q}_b 
\rangle_{\op{\mathcal{D}}}$, together with the anticommutation relations (\ref{eq:59}) for the 
set $\bigl\{ \op{\gamma}_{l}, \gopp{\gamma}{l} \bigr\} $. For example, writing
$\op{Q}_a$ and $\op{Q}_b$ in terms of the ket $\ket{Q_a}$ and the bra 
$\bra{\bar{Q}_b}$, respectively, one has
\begin{eqnarray}\label{eq:79}
\langle \op{Q}_a \op{Q}_b \rangle_{\op{\mathcal{D}}} &=\sum_{l=1}^{d}
\frac{\bigl[\op{\gamma}_l,\op{Q}_b  \bigr]_+}{1+\rme^{-\vartheta_l}} \braket{\tilde{\gamma}'_l}{Q_a} 
+\sum_{l=1}^{d}\frac{\bigl[\gopp{\gamma}{l},\op{Q}_b  \bigr]_+}{1+\rme^{\vartheta_l}} 
\braket{\bar{\gamma}_l}{Q_a} \nonumber\\
&=\sum_{l=1}^{d}  \frac{\braket{\bar{Q}_b}{\gamma_l} 
\braket{\tilde{\gamma}'_l}{Q_a}}{1+\rme^{-\vartheta_l}} + \sum_{l=1}^{d}
\frac{\braket{\bar{Q}_b}{\bar{\tilde{\gamma}}'_l} 
\braket{\bar{\gamma}_l}{Q_a}}{1+\rme^{\vartheta_l}} .
\end{eqnarray}
Equivalently, with the bra $\bra{\bar{Q}_a}$ and the ket $\ket{Q_b}$, one gets
\begin{equation}\label{eq:80}
\langle \op{Q}_a \op{Q}_b \rangle_{\op{\mathcal{D}}} = 
\sum_{l=1}^{d}  \frac{\braket{\bar{Q}_a}{\gamma_l}
  \braket{\tilde{\gamma}'_l}{Q_b}}{1+\rme^{\vartheta_l}} + \sum_{l=1}^{d}
\frac{\braket{\bar{Q}_a}{\bar{\tilde{\gamma}}'_l}
  \braket{\bar{\gamma}_l}{Q_b}}{1+\rme^{-\vartheta_l}} .
\end{equation}
Taking into account the completeness relation (\ref{eq:77}) in the extended
one-body space, the two expressions (\ref{eq:79}-\ref{eq:80}) are
identical. Besides, they have a perfectly well defined limit when
$\{\vartheta_l \to \infty \}$, where the operator $\op{\mathcal{D}}$
identifies to the dyad $\ket{\Phi} \bra{\tilde{\Phi}'}$, and gives access to
the matrix elements between the two vacua (HF or HFB). Irrespective of the
factors $\op{Q}_a$ considered, the contractions $\langle \op{Q}_a \op{Q}_b
\rangle_{\op{\mathcal{D}}}$ can also be deduced from those between two
elementary fermionic operators that define the generalized one-body density
matrix $\mathcal{R}$
\begin{equation}
\mathcal{R} \!=\! \left(\begin{array}{cc} \rho & \kappa\\
\tilde{\kappa} & \tilde{\rho}
\end{array}\right) \,\,\, \mbox{with}\,\,\,  \left\{ 
\begin{array}{c} \rho_{{i}, {{j}}} = 
\langle  \op{c}^{\dagger}_{{j}} 
\op{c}^{\phantom{\dagger}}_{{i}} \rangle_{\op{\mathcal{D}}} \\[.4em]
  \tilde{\rho} _{{i}, {{j}}} = 
\langle  \op{c}^{\phantom{\dagger}}_{{j}} 
\op{c}^{\dagger}_{{i}} \rangle_{\op{\mathcal{D}}}
\end{array}\right. \,\,\, \mbox{and}\,\,\,  \left\{ 
\begin{array}{c} \kappa_{{i}, {{j}}} = 
\langle  \op{c}^{\phantom{\dagger}}_{{j}} 
\op{c}^{\phantom{\dagger}}_{{i}} \rangle_{\op{\mathcal{D}}} \\[.4em]
\tilde{\kappa} _{{i}, {{j}}} = 
\langle  \op{c}^{\dagger}_{{j}} 
\op{c}^{\dagger}_{{i}} \rangle_{\op{\mathcal{D}}}
\end{array}\right. .
\end{equation}
In other words, $\aver{\varpi {i}}{\mathcal{R}}{\varpi' {j}} = 
\langle \op{c}^{\dagger}_{\varpi' {j}} 
\op{c}^{\phantom{\dagger}}_{\varpi {i}} \rangle_{\op{\mathcal{D}}}$ and thus
\begin{eqnarray}
\langle \op{Q}_a \op{Q}_b \rangle_{\op{\mathcal{D}}} &= \sum_{\varpi {i}}
\sum_{\varpi' {j}} \braket{\bar{Q}_b}{\varpi {i}} \langle \op{c}^{\dagger}_{\varpi' {j}} 
\op{c}^{\phantom{\dagger}}_{\varpi {i}} \rangle_{\op{\mathcal{D}}}
\braket{\varpi' {j}}{Q_a} = \aver{{\bar{Q}_b}}{\mathcal{R}}{Q_a} \nonumber\\
&=\sum_{\varpi {i}} \sum_{\varpi' {j}} \braket{\bar{Q}_a}{\varpi {i}}
\langle \op{c}^{\phantom{\dagger}}_{\varpi {i}} \op{c}^{\dagger}_{\varpi' {j}} 
\rangle_{\op{\mathcal{D}}}\braket{\varpi' {j}}{Q_b} = \aver{{\bar{Q}_a}}{\unitmatrix-\mathcal{R}}{Q_b}\,.
\end{eqnarray}
By virtue of the results (\ref{eq:79}-\ref{eq:80}) coming from the
demonstration of Wick's theorem, $\mathcal{R}$ is thus given, in the limit of
zero temperature $\{\vartheta_l \to \infty \}$, by
\begin{eqnarray}
\mathcal{R}&=&\sum_{l=1}^{d}\ket{\gamma_l}\bra{\tilde{\gamma}'_l} = 
\sum_{l,m=1}^{d}\ket{\gamma_l} \left(\mathbb{F}^{-1} \right)_{lm}\bra{\gamma'_m},\nonumber\\
\unitmatrix&-&\mathcal{R}=\sum_{l=1}^{d}\ket{\bar{\tilde{\gamma}}'_l}\bra{\bar{\gamma}_l}= 
\sum_{l,m=1}^{d}\ket{\bar{\gamma}'_l} \left(\mathbb{F}^{-1} \right)_{ml}
\bra{\bar{\gamma}_m} .
\end{eqnarray}
Eventually, $\langle \op{Q}_1 \op{Q}_2 \ldots \op{Q}_M
\rangle_{\op{\mathcal{D}}}$ is a functional $\mathcal{Q} [\mathcal{R}]$ of the
reduced density matrix $\mathcal{R}$ which results from the repeated
application of the recursive algorithm (\ref{eq:74}) to estimate the
expectation values of $M-2$, and then $M-4$, $\ldots$ factors. As a matter of
fact, $\mathcal{Q} [\mathcal{R}]$ can be directly obtained by noting that this
recurrence relation is exactly that of the development of a Pfaffian, that is
to say of the square root of the determinant of an antisymmetric matrix
\cite{Cay52, Baj08}. Let us introduce such a matrix $\mathcal{C}$ of dimension
$M\times M$ with upper triangular entries given by the binary contractions of
factors $\op{Q}_a$ of the considered product $\mathcal{C}_{a,b} = 
\langle \op{Q}_a \op{Q}_b \rangle_{\op{\mathcal{D}}}$, $a<b \in \{1,\ldots,
M\}$. The Pfaffian $\mbox{Pf}(\mathcal{C})$ can then be found according to a
procedure similar to that of calculating a determinant, i.e., through the
expansion, for example, along the first row \cite{Baj08}
\begin{equation}\label{eq:84}
\mbox{Pf}(\mathcal{C}) = \sum_{b=2}^{M} (-1)^b \mathcal{C}_{1,b} 
\mbox{Pf}(\mathcal{C}^{(1,b)}),
\end{equation}
where $\mathcal{C}^{(1,b)}$ is the sub-matrix obtained by removing the first
row  and the $b$-th column. We thus immediately obtain the identity 
$\mathcal{Q} [\mathcal{R}] = \mbox{Pf}(\mathcal{C})$ by mathematical
induction: For two factors, the definitions of the matrix $\mathcal{C}$ and of
the Pfaffian indeed lead to
\begin{equation}
\langle \op{Q}_1 \op{Q}_2 \rangle_{\op{\mathcal{D}}} = \mathcal{C}_{1,2} =
\mbox{Pf} \left(\begin{array}{cc} 0&\mathcal{C}_{1,2}\\-\mathcal{C}_{1,2}&0
\end{array}\right) .
\end{equation}
Assuming this result to be valid for $M-2$ factors, 
$\langle \op{Q}_2 \ldots \op{Q}_{b-1} \op{Q}_{b+1}\ldots \op{Q}_M
\rangle_{\op{\mathcal{D}}} = \mbox{Pf}(\mathcal{C}^{(1,b)})$ follows, so that the
relations (\ref{eq:74}, \ref{eq:84}) complete the proof by leading to 
\begin{equation}
\langle \op{Q}_1 \op{Q}_{2} \ldots \op{Q}_M
\rangle_{\op{\mathcal{D}}} = \mbox{Pf}(\mathcal{C}) . 
\end{equation}
To our knowledge, this connection between the Wick theorem and the Pfaffians
has been originally highlighted by E.~Lieb~\cite{Lie68}, following 
M.~Gaudin's work \cite{Gau60}. It allows, \textit{via} the explicit form of the Pfaffian of a matrix in 
terms of its elements, to synthesize the previous results in the form
\begin{eqnarray}\label{eq:85}
\langle \op{Q}_1 \op{Q}_2  \ldots \op{Q}_M
\rangle_{\op{\mathcal{D}}} = \mathcal{Q}[\mathcal{R}] =
\mbox{ Pf}(\mathcal{C}) = \sum_{\pi} \varepsilon_{\pi}
\mathcal{C}_{\pi(1),\pi(2)} \ldots \mathcal{C}_{\pi(M-1),\pi(M)} , \nonumber\\
\mathcal{C}_{ab}=\langle \op{Q}_a \op{Q}_b \rangle_{\op{\mathcal{D}}} =
\aver{\bar{Q}_b}{\mathcal{R}}{Q_a} =\aver{\bar{Q}_a}{\unitmatrix-\mathcal{R}}{Q_b}\,.
\end{eqnarray}
Here, recalling that $M$ is even, the sum is performed on the $(M-1)!!$
permutations $\pi$ of the set 
$\{1,2,\ldots ,M\}$ satisfying the constraints $\pi(2r-1)<\pi(2r+1)$
($r=1,\ldots ,M/2-1$) and $\pi(2r-1)<\pi(2r)$ ($r=1,\ldots ,M/2)$, with $
\varepsilon_{\pi}$  designating their signature. On top of the formal aspects,
the reformulation of the Wick theorem as a Pfaffian is particularly well
suited for numerical implementations for a large number of factors $\op{Q}_a$,
thanks to effective numerical methods to evaluate $\mbox{Pf}(\mathcal{C})$
through the determination of a tridiagonal antisymmetric form for the matrix
$\mathcal{C}$ \cite{Wim12}.

Finally, the case $\op{Q}_1 \op{Q}_2 \ldots \op{Q}_M$ that we have not treated
yet, where the product involves an odd number of factors, is in fact trivial
and systematically leads to zero matrix elements. Indeed, the expression
(\ref{eq:61}) of the vector $\rket{n_1 \ldots n_d}$, as 
well as the canonical anticommutation relations (\ref{eq:59}) satisfied by
the operators $\gopp{\gamma}{l}$ and $\op{\gamma}_{l}$, show that these
operators increase and decrease by one the occupation number $n_l$,
respectively. An odd number of factors $\op{Q}_a$ cannot therefore
keep the total number of excitations $n_1+ \ldots +n_d$, while the Gaussian
ansatz $\op{\mathcal{D}}$ preserves it. As a consequence, 
\begin{equation}
\mbox{Tr}[\op{Q}_1 \op{Q}_2  \ldots
\op{Q}_M \op{\mathcal{D}}] = \sum_{\{n_1 \ldots n_d\}} \lbra{n_1 \ldots n_d} 
\op{\mathcal{D}} \op{Q}_1 \ldots \op{Q}_M \rket{n_1 \ldots n_d},
\end{equation}
vanishes necessarily. 

\subsection{Overlaps}\label{sec:Overlaps}

Let us now show that Wick's theorem does also give access to the overlaps
$\braket{\Phi'}{\Phi}$, which are necessary to determine the matrix elements
(\ref{eq:54}) between two HF or HFB wavefunctions. With at least one HFB
state among $\ket{\Phi}$, $\ket{\Phi'}$ it should be noted that only the
modulus of the overlap $\braket{\Phi'}{\Phi}$ has been determined so far
through Onishi's formula (\ref{eq:56}). In every approach based on a
superposition of such wavefunctions, the phase of $\braket{\Phi'}{\Phi}$
obviously plays a key role and a procedure for determining it was proposed \textit{via}
the spectrum of the non-hermitian matrix $\mathbb{F}$ \cite{Nee82}. However,
this method remains numerically expensive and, as a consequence, has only been
concretely used for problems characterized by one-body spaces of small
dimension \cite{Sch04}. The alternative use of Pfaffians, to directly access the
overlap between two HFB states, was initiated in 2009 by Robledo through a
calculation using Grassmann variables \cite{Rob09}, which was subsequently
resumed in terms of a process similar to the one that we will follow \cite{Ber12}. 

The idea rests upon Wick's theorem, formulated in terms of a Pfaffian (\ref{eq:85}), after noting 
that all developments carried out for its
demonstration remain valid if the expectations values are calculated in the
vacuum of fermions $\ket{\;}$. Moreover, irrespective of the HF or HFB nature
of each of the two wavefunctions $\ket{\Phi}$, $\ket{\Phi'}$, they can be
written as a product of factors that linearly depend on creation
($\op{c}^{\dagger}_{{i}}$) and annihilation 
($\op{c}^{\vphantom{\dagger}}_{{i}}$) operators. Consequently, their
overlap $\braket{\Phi'}{\Phi}$ identifies to the expectation value in the
vacuum $\ket{\;}$ of such products, and it may therefore be determined thanks
to Wick's theorem. As a simple example, let us first consider the case of two
Slater determinants $\ket{\Phi} = \op{c}^{\dagger}_{\phi_1} \ldots
\op{c}^{\dagger}_{\phi_N}\ket{\;}$ and $\ket{\Phi'} = \op{c}^{\dagger}_{\phi'_1}
\ldots \op{c}^{\dagger}_{\phi'_N}\ket{\;}$, which overlap is easily
obtained through a direct calculation: $\braket{\Phi'}{\Phi} =
\mbox{det}(\phi^{'\dagger} \phi)$. Here $\phi$ and $\phi'$ are rectangular
tables of dimensions $d\times N$ , defined by the components 
$\{\phi_{{i},n}\} $ and $\{\phi'_{{i},n}\}$ of the
occupied individual states of $\ket{\Phi}$ and  $\ket{\Phi'}$,
respectively. Noting that $\bra{\Phi'} = (-1)^{N(N-1)/2} \bra{\;} 
\op{c}^{\vphantom{\dagger}}_{\phi'_1} \ldots
\op{c}^{\vphantom{\dagger}}_{\phi'_N}$, 
Wick's theorem leads to
$\braket{\Phi'}{\Phi} = (-1)^{N(N-1)/2} \aver{\;} {\op{c}^{\vphantom{\dagger}}_{\phi'_1} \ldots  
\op{c}^{\vphantom{\dagger}}_{\phi'_N}\op{c}^{\dagger}_{\phi_1} \ldots
\op{c}^{\dagger}_{\phi_N}}{\;} = (-1)^{N(N-1)/2} \mbox{Pf}(\mathcal{C}) $
where $\mathcal{C}$ is the antisymmetrized matrix of the binary
contractions. Here, it is thus a $2N\times 2N$ matrix which elements are given by  
\begin{equation}
\fl\mathcal{C}_{{{n}}{{p}}} = \left\{ \begin{array}{ll}
\aver{\;}{\op{c}^{\vphantom{\dagger}}_{\phi'_{{{n}}}}\op{c}^{\vphantom{\dagger}}_{\phi'_{{{p}}}}}{\;
}
= 0 & \mbox{if   } 1\leq {{n}}<{{p}}\leq N \\
\aver{\;}{\op{c}^{\vphantom{\dagger}}_{\phi'_{{{n}}}}\op{c}^{\dagger}_{\phi_{{{p}}-N}}}{\;}
= \braket{\phi'_{{{n}}}}{\phi_{{{p}}-N}}=\bigl( \phi^{'\dagger}\phi
\bigr)_{{{n}},{{p}}-N} & \mbox{if} \left\{\!\! \begin{array}{l}
1\leq {{n}} \leq N \\ 
N+1 \leq {{p}} \leq 2N \end{array} 
\right. \\
\aver{\;}{\op{c}^{\dagger}_{\phi_{{{n}}-N}}\op{c}^{\dagger}_{\phi_{{{p}}-N}}}{\;}
= 0 & \mbox{if  } N+1 \leq {{n}} < {{p}} \leq 2N \end{array} \right..
\end{equation}
In other words, 
\begin{equation}
\mbox{Pf}(\mathcal{C}) =  \mbox{Pf}\left(
\begin{array}{cc} 0_{N\times N}&\phi^{'\dagger}\phi\\-
\phi^{T}\phi^{'*}&0_{N\times N}
\end{array}\right) = (-1)^{N(N-1)/2} \mbox{det}\bigl(\phi^{'\dagger}\phi\bigr).
\end{equation}
This follows from the properties of the Pfaffian \cite{Baj08} and we therefore
find the expected expression for the overlap $\braket{\Phi'}{\Phi}$ between
two HF vacua. The calculation is in all respects similar for two normalized
HFB states, when expressed in terms of their respective quasiparticles
$\{\op{\gamma}_n\}$ and $\{\op{\gamma}'_n\}$ under the form of (\ref{EQ:10})
\begin{equation}\label{eQ:109}
\ket{\Phi} = \frac{1}{\nu_1 \ldots \nu_{d/2}} \op{\gamma}_1 \ldots
  \op{\gamma}_d  \ket{\;},\;\;\; \ket{\Phi'} = \frac{1}{\nu'_1
  \ldots \nu'_{d/2}} \op{\gamma}'_1 \ldots \op{\gamma}'_d  \ket{\;}\,.
\end{equation}
We denote by $\{ U_{{i},n},  V_{{i},n} \}$  ($\{ U'_{{i},n},  V'_{{i},n} \}$)
the amplitudes of the Bogoliubov transformation associated to $\ket{\Phi}$
($\ket{\Phi'}$). In equation (\ref{eQ:109}), the set of real numbers $\{
v_{\alpha}\}$ ($\{v'_{\alpha}\}$) define the Bloch-Messiah-Zumino
decomposition of the matrix $V$ ($V'$). 
Wick's theorem then allows to express the
expectation value $\aver{\;}{\op{\gamma}^{'\dagger}_1 \ldots
  \op{\gamma}^{'\dagger}_d \op{\gamma}^{\vphantom{\dagger}}_{1} \ldots
  \op{\gamma}^{\vphantom{\dagger}}_{d}}{\;}$ of the product of all
quasiparticle operators in terms of binary contractions 
\begin{equation}
\mathcal{C}_{{{n}}{{p}}} = \left\{ \begin{array}{ll}
\aver{\;}{\op{\gamma}^{'\dagger}_{{n}} \op{\gamma}^{'\dagger}_{{p}}}{\;}  
=\bigl(V^{'T}U'\bigr)_{{{n}}{{p}}} & \mbox{if  } 1\leq {{n}}<{{p}}\leq d \\
\aver{\;}{\op{\gamma}^{'\dagger}_{{n}} \op{\gamma}_{{{p}}-d} }{\;}
= \bigl(V^{'T}V^*\bigr)_{{{n}},{{p}}-d} & \mbox{if} \left\{\!\! \begin{array}{l}
1\leq {{n}} \leq d \\ 
d+1 \leq {{p}} \leq 2d \end{array}\right. \\
\aver{\;}{\op{\gamma}_{{{n}}-d} \op{\gamma}_{{{p}}-d} }{\;}
= \bigl(U^{\dagger}V^*\bigr)_{{{n}}-d,{{p}}-d} & \mbox{if  } d+1 \leq {{n}} < {{p}} \leq 2d 
\end{array}\right.,
\end{equation}
since $(V^{'T}U')_{{{n}}{{p}}} = \sum_{{i},{j}} 
V'_{{j},{{n}}} U'_{{i},{{p}}}
\aver{\;}{\op{c}^{\vphantom{\dagger}}_{{j}}\op{c}^{\dagger}_{{i}}}{\;} $, and
accordingly for the other two types of matrix elements.
Noting that the unitarity of Bogoliubov's transformations implies that the
matrices $V^{'T}U'$  and $U^{\dagger}V^*$ are antisymmetric, the overlap
between two HFB states finally reads
\begin{equation}
\braket{\Phi'}{\Phi} =  \frac{(-1)^{d(d-1)/2}}
{\nu'_1\nu^{\vphantom{'}}_1 \ldots \nu'_{d/2}\nu^{\vphantom{'}}_{d/2}} \mbox{Pf}\left(
\begin{array}{cc} V^{'T}U'&V^{'T}V^*\\
-V^{\dagger}V'&U^{\dagger}V^*
\end{array}\right) .
\end{equation}

We refer to \cite{Ber12} and \cite{Rob09} to prove that this expression
reduces to Onishi's formula (\ref{eq:56}) for the square of the modulus of
the scalar product between the two wavefunctions. Finally, in the hybrid case
of a Bogoliubov vacuum $\ket{\Phi}$ (see equation (\ref{EQ:10})) and a Slater determinant
$\ket{\Phi'} = \op{c}^{\dagger}_{\phi'_1} \ldots \op{c}^{\dagger}_{\phi'_N}
\ket{\;}$, one now needs to calculate 
$\aver{\;}{\op{c}^{\vphantom{\dagger}}_{\phi'_1} \ldots
  \op{c}^{\vphantom{\dagger}}_{\phi'_N}\op{\gamma}_1 \ldots
  \op{\gamma}_d}{\;}$, i.~e., the contractions 
\begin{equation}
\fl\mathcal{C}_{{{n}}{{p}}} = \left\{ \begin{array}{ll}
\aver{\;}{\op{c}^{\vphantom{\dagger}}_{\phi'_{{{n}}}}\op{c}^{\vphantom{\dagger}}_{\phi'_{{{p}}}}}{\;
}
= 0 & \mbox{if  }1\leq {{n}}<{{p}}\leq N \\
\aver{\;}{\op{c}^{\vphantom{\dagger}}_{\phi'_{{{n}}}}\op{\gamma}^{\vphantom{\dagger}}_{{{p}}-N}}{\;}
=
\bigl(\phi^{'\dagger}V^*\bigr)_{{{n}},{{p}}-N} & \mbox{if} \left\{\!\! \begin{array}{l} 
1\leq {{n}} \leq N \\ 
N+1 \leq {{p}} \leq N+d \end{array} \right. \\
\aver{\;}{\op{\gamma}^{\vphantom{\dagger}}_{{{n}}-d}\op{\gamma}^{\vphantom{\dagger}}_{{{p}}-N}}{\;}
= \bigl(U^{\dagger}V^*\bigr)_{{{n}}-N,{{p}}-N} = 0 & \mbox{if  } N+1 \leq {{n}} < {{p}} \leq
N+d , \end{array} \right.
\end{equation}
where we used $ \bigl(\phi^{'\dagger}V^*\bigr)_{{{n}},{{p}}-N} = \sum_{{i},{j}} 
\phi^{'*}_{{j},{{n}}} V^*_{{i},{{p}}-N}
\aver{\;}{\op{c}^{\dagger}_{{i}}\op{c}^{\vphantom{\dagger}}_{{j}}} {\;} $. 
As a result, the overlap is now given by the Pfaffian of a square matrix of dimensions $N+d$
\begin{equation}
\braket{\Phi'}{\Phi} = 
\frac{(-1)^{N(N-1)/2} }{\nu^{\vphantom{'}}_1 \ldots \nu^{\vphantom{'}}_{d/2}} 
\mbox{Pf}\left(
\begin{array}{cc} 0_{N\times N}&\phi^{'\dagger}V^*\\
-V^{\dagger}\phi^{'*}&U^{\dagger}V^*
\end{array}\right) \,. 
\end{equation}

\section{Numerical illustration with the Hubbard model}\label{sec:numex}

Simultaneously introduced in 1963 by J. Hubbard \cite{Hub63}, M. C. Gutzwiller
\cite{Gut63} et J. Kanamori \cite{Kan63}, the Hubbard model is among the
simplest and the most commonly used ones in theoretical condensed-matter
physics. It aims to grasp the generic properties of spin-1/2 fermions moving
on a lattice by hopping between neighboring sites $<{\bf r}, {\bf r'}>$ and
experiencing a local two-body interaction of strength $U$. In second-quantized
form, the Hamiltonian is given by
\begin{equation}\label{EQ:114}
\op{H} = -t \sum_{<{\bf r}, {\bf r'}>, \sigma = \uparrow, \downarrow }
\op{c}^{\dagger}_{{\bf r}\sigma}
\op{c}^{\vphantom{dagger}}_{{\bf r'}\sigma} + U \sum_{{\bf r}}
\op{n}^{\vphantom{dagger}}_{{\bf r}\uparrow} 
\op{n}^{\vphantom{dagger}}_{{\bf r}\downarrow} ,
\end{equation}
with $t$ the hopping integral; The fermionic creation, annihilation and density
operators at site ${\bf r}$ with spin $\sigma \in \{\uparrow, \downarrow\}$
are $\op{c}^{\dagger}_{{\bf r}\sigma}$, 
$\op{c}^{\vphantom{dagger}}_{{\bf r}\sigma}$, and 
$\op{n}^{\vphantom{dagger}}_{{\bf r}\sigma} = \op{c}^{\dagger}_{{\bf r}\sigma} 
\op{c}^{\vphantom{dagger}}_{{\bf r}\sigma}$, respectively. In the positive $U$
regime, the on-site repulsion stands for a perfectly screened Coulomb
interaction and the model received a considerable renewed interest in
two-dimensional (2D) geometry after Anderson's proposal in connection to
high-$T_c$ SC cuprates \cite{And87}. However, there is still no consensus
about the 
adequacy of the repulsive 2D Hubbard model to capture the interplay between
$d$-wave superconductivity, magnetism and inhomogeneous phases of copper
oxides. In particular, constrained-path auxiliary-field QMC simulations do not
give a clear answer as to the relevance, or not, of $d$-wave pair condensation
that is yet obtained with variational schemes \cite{Zha97b,Gue99}.

As an application of the above developed ``Phaseless QMC'' approach, we focus
here on the attractive sector, for which the stochastically explored HFB
states \textit{a priori} constitute an appealing approximation. Moreover, one
only has to consider spin polarized systems at half-filling for the
ground-state correlations to be directly related to those exhibited in the
doped repulsive case \cite{Shi1}. This result is an immediate consequence of
Shiba's particle-hole transformation \cite{Shiba}, given for a 2D square lattice by
\begin{equation}
\op{c}^{\dagger}_{{\bf r} \uparrow} \to
\op{\underline{c}}^{\dagger}_{{\bf r} \uparrow},\;\;\;
\op{c}^{\dagger}_{{\bf r} \downarrow} \to (-1)^{x+y} \op{\underline{c}}^{\vphantom{\dagger}}_{{\bf 
r} \downarrow}.
\end{equation}
Indeed, up to an additive constant, the Hubbard Hamiltonian is recovered
for the transformed fermions but with the sign of the on-site interaction 
changed into its opposite. Moreover, asymmetrical fillings of the two spin
projections $N_{\uparrow} = \mathcal{N} (1-\delta)/2$, 
$N_{\downarrow} = \mathcal{N} (1+\delta)/2$ (where $\mathcal{N}$ is the number
of sites and $0\leq \delta \leq 1$)  become 
$\underline{N}_{\uparrow} = \underline{N}_{\downarrow} =
\mathcal{N} (1-\delta)/2$. After transformation they thus correspond to a
hole doping $\delta$. In addition, an SC homogeneous phase is linked to an
antiferromagnetic order which is relevant for the repulsive model in the
vicinity of the Mott insulator: $\langle \op{c}^{\vphantom{\dagger}}_{{\bf r} \downarrow}
\op{c}^{\vphantom{\dagger}}_{{\bf r} \uparrow} \rangle  \to (-1)^{x+y} 
\langle \op{\underline{c}}^{\dagger}_{{\bf r} \downarrow}
\op{\underline{c}}^{\vphantom{\dagger}}_{{\bf r} \uparrow} \rangle$. 
Likewise, a $d$-wave spin-density wave is the counterpart of the
superconductivity expected for $U>0$, $\delta>0$: 
$(-1)^{x+y} \langle \op{c}^{\dagger}_{{\bf r} \downarrow}
\op{c}^{\vphantom{\dagger}}_{{\bf r}+{\bf u}_x \uparrow} -
\op{c}^{\dagger}_{{\bf r} \downarrow}
\op{c}^{\vphantom{\dagger}}_{{\bf r}+{\bf u}_y \uparrow} \rangle \to
\langle \op{\underline{c}}^{\vphantom{\dagger}}_{{\bf r} \downarrow}
\op{\underline{c}}^{\vphantom{\dagger}}_{{\bf r}+{\bf u}_x  \uparrow} -
\op{\underline{c}}^{\vphantom{\dagger}}_{{\bf r} \downarrow}
\op{\underline{c}}^{\vphantom{\dagger}}_{{\bf r}+{\bf u}_y  \uparrow}\rangle$. 
In the attractive and spin-polarized regime that we
consider, these observations thus motivate the construction of a HFB
approximation from a simple one-body Hamiltonian $\op{h}_0$, including the two
previous channels,
\begin{eqnarray}\label{EQ:119}
\op{h}_0 = &-&t \sum_{<{\bf r}, {\bf r'}>, \sigma = \uparrow, \downarrow }
\op{c}^{\dagger}_{{\bf r}\sigma}
\op{c}^{\vphantom{dagger}}_{{\bf r'}\sigma} + 
\Delta \sum_{{\bf r}} \left( \op{c}^{\dagger}_{{\bf r} \uparrow} 
\op{c}^{\dagger}_{{\bf r} \downarrow} + \op{c}^{\vphantom{\dagger}}_{{\bf r} \downarrow}
\op{c}^{\vphantom{\dagger}}_{{\bf r} \uparrow} \right) \nonumber\\
&+& m_d \sum_{{\bf r}, {\bf l} \in \{ \pm{\bf u}_x, \pm{\bf u}_y \} }
f({\bf l}) (-1)^{x+y} \left( \op{c}^{\dagger}_{{\bf r} \uparrow}
\op{c}^{\vphantom{\dagger}}_{{\bf r}+{\bf l} \downarrow} + 
\op{c}^{\dagger}_{{\bf r} \downarrow}
\op{c}^{\vphantom{\dagger}}_{{\bf r}-{\bf l} \uparrow} \right),
\end{eqnarray}
where $f(\pm{\bf u}_x) = 1$, $f(\pm{\bf u}_y) = -1$. Here, $\Delta$ and
$m_d$ play the role of the parameters related to the orders respectively
associated to the condensation of Cooper pairs with $s$-wave symmetry and 
$d$-wave bond-spin antiferromagnetism. They are here \textit{a priori} fixed,
and no self-consistency is considered. In other words, we limit ourselves to
the determination of the HFB ground state $\ket{\Phi_0}$ of (\ref{EQ:119})
under the constraint that both spin sectors are correctly populated on
average. In the following, the stochastic dynamics at the heart of the
``Phaseless QMC'' scheme will be initiated by this vector $\ket{\Phi_0}$ and
guided by a trial state $\ket{\Psi_T}$ stemming from its projection on the
considered fermionic numbers $N_{\uparrow}$, $N_{\downarrow}$, i.e.
\begin{equation}\label{EQ:120}
\ket{\Psi_T} = \op{P}_{N_{\uparrow}, N_{\downarrow}}\ket{\Phi_0} .
\end{equation}
It should be noted that the presence of the projector $\op{P}_{N_{\uparrow},
  N_{\downarrow}}$ is essential to ensure a strict preservation of the total
density as well as the spin polarization in the QMC simulation: With the
choice (\ref{EQ:120}) for the trial state, 
$\langle \op{N}_{\uparrow}\rangle_{\Psi_T,\Phi_{\tau}}$ and 
$\langle \op{N}_{\downarrow}\rangle_{\Psi_T,\Phi_{\tau}}$ remain unchanged irrespective
of the HFB stochastic realization $\ket{\Phi_{\tau}}$ and independently of the
imaginary time $\tau$. In practice, the restoration of the quantum numbers
$(N_{\uparrow}, N_{\downarrow})$ is carried out by the superposition of gauge
transformations
\begin{equation}\label{EQ:121}
\op{P}_{N_{\uparrow}, N_{\downarrow}} = \frac{1}{4 \pi^2} \int_0^{2 \pi}\!\!\!
\int_0^{2 \pi} \! \prod_{\sigma} d\varphi_{\sigma} \rme^{-i \varphi_{\sigma} N_{\sigma}} \rme^{\rmi 
\varphi_{\sigma} \op{N}_{\sigma}} \,.
\end{equation}
Given the developments presented in Section~\ref{sec:sto}, each of them
transforms the HFB wavefunction $\ket{\Phi_0}$ into another one 
$\ket{\Phi_0'(\bm{\varphi})}$
where the vector $\bm{\varphi}$ gathers the two gauge angles 
$(\varphi_{\uparrow},\varphi_{\downarrow})$. In the extended one-body space,
the states $\ket{\gamma_n}_0$ and $\ket{\gamma_n'(\bm{\varphi})}_0$ of their
respective quasiparticles are related by
\begin{equation}\label{EQ:122}
\ket{\gamma_n'(\bm{\varphi})}_0 = 
\left(
\begin{array}{cc} 
\rme^{\rmi (\varphi_{\uparrow} + \varphi_{\downarrow})} \unitmatrix_{d\times d}&0_{d\times d}\\
0_{d\times d}&\rme^{-i (\varphi_{\uparrow} + \varphi_{\downarrow})} \unitmatrix_{d\times d}
\end{array}\right) \ket{\gamma_n}_0 \,.
\end{equation}
Thus, the trial state $\ket{\Psi_T}$ appears as a linear combination of
Bogoliubov vacua so that any local estimator $\langle
\op{A}\rangle_{\Psi_T,\Phi_{\tau}}$ is easily evaluated through the overlaps 
$\braket{\Phi_0'(\bm{\varphi})}{\Phi_{\tau}}$ and the extended Wick theorem
which gives access to 
$\aver{\Phi_0'(\bm{\varphi})}{\op{A}}{\Phi_{\tau}}/\braket{\Phi_0'(
    \bm{\varphi})}{\Phi_{\tau}}$. Finally, the implementation of the
``Phaseless QMC'' approach to the Hubbard model requires for the Hamiltonian (\ref{EQ:114}) a 
quadratic form of general one-body operators, thus
ensuring that the Bogoliubov transformation matrices are not real throughout
the stochastic evolution. In this case, this step is immediate by writing
\begin{equation}
\op{H} = -t \sum_{<{\bf r}, {\bf r'}> \\ \sigma = \uparrow, \downarrow }
\op{c}^{\dagger}_{{\bf r}\sigma}
\op{c}^{\vphantom{dagger}}_{{\bf r'}\sigma} + \frac{U}{4} \sum_{{\bf r}}
\Bigl[\bigl( \op{n}^{\vphantom{dagger}}_{{\bf r}\uparrow} +
\op{n}^{\vphantom{dagger}}_{{\bf r}\downarrow} 
\bigr)^2 -\bigl( \op{n}^{\vphantom{dagger}}_{{\bf r}\uparrow} -
\op{n}^{\vphantom{dagger}}_{{\bf r}\downarrow} 
\bigr)^2\Bigr]\,.
\end{equation}
With $U<0$, the introduction of local spin polarization 
$\op{n}^{\vphantom{dagger}}_{{\bf r}\uparrow} -
\op{n}^{\vphantom{dagger}}_{{\bf r}\downarrow} $ indeed leads to purely
imaginary fluctuating contributions in the Brownian motion (\ref{EQ:27})
of the quasiparticles, while the on-site density
$\op{n}^{\vphantom{dagger}}_{{\bf r}\uparrow} +
\op{n}^{\vphantom{dagger}}_{{\bf r}\downarrow} $ induces a strictly real
diffusive part. 
\begin{figure}[t!]
\psfrag{XXXXX}{$\,\,\,\,\bm{\bar{E}_{\tau}\,[t]}$}
\psfrag{YYYY}{$\,\,\,\,\bm{\tau\,[1/t]}$}
\begin{center}
\includegraphics*[width=\linewidth]{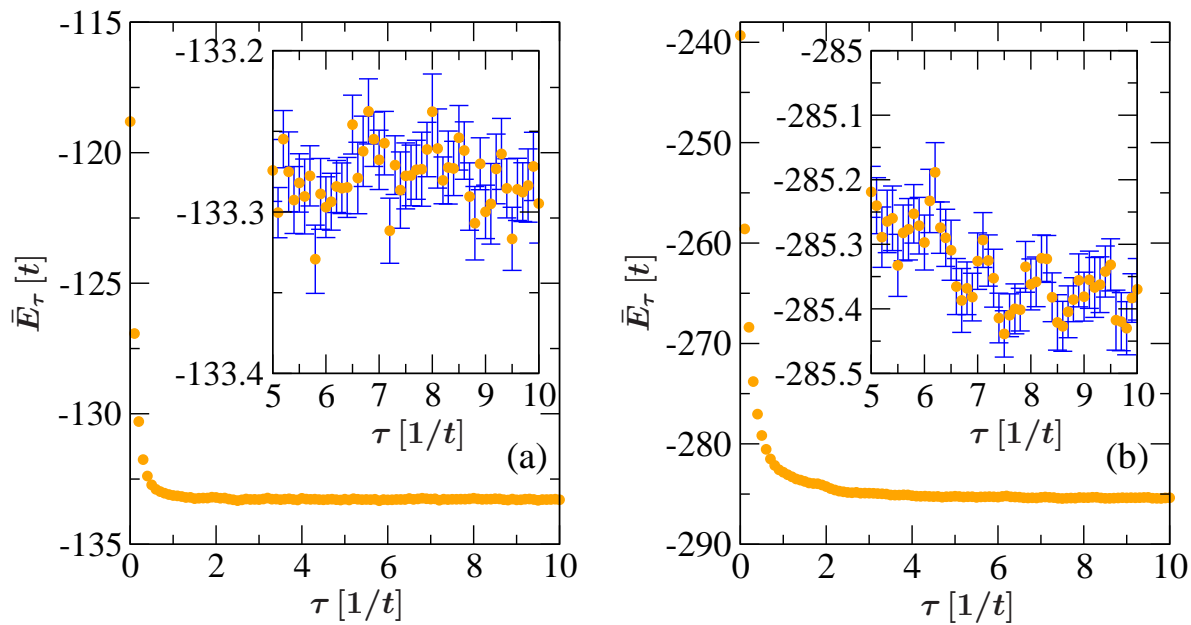}
\end{center}
\caption{(Color online) 
Imaginary-time evolution of the energy $\bar{E}_{\tau}$, the average over all
populations of 
$E_{\tau} = \mathbb{E}_{\tilde{\Pi}}\bigl[\Re{\bigl( \langle \op{H} 
\rangle_{\Psi_T,\Phi_{\tau}}\bigr)}\bigr]$, for the
Hubbard Model with $U=-8 t$.
In panel (a), a half-filled $6 \times 6$ cluster
with $N_{\uparrow} = 12$  and $N_{\downarrow} = 24$ (using periodic boundary
conditions in both directions) is considered. Panel (b) addresses the
half-filled $8 \times 8$ cluster with a smaller spin polarization
$N_{\uparrow} = 31$  and $N_{\downarrow} = 33$ (using mixed
periodic-antiperiodic boundary conditions, to ease the comparison with
existing variational calculations in the repulsive model). In both cases the
trial state follows from the constrained diagonalization of the one-body
Hamiltonian $\op{h}_0$ (\ref{EQ:119}) with the fixed order parameters
$\Delta = 0.5 t$ and $m_d=0.1 t$. The averages and error bars arise from 25
and 40 independent populations of $N_w=1000$ walkers in part (a) and (b),
respectively. The statistical fluctuations are smaller than the size of the
points when the full evolution is presented. Therefore, they are made visible
in the corresponding insets.}
\label{fig:energy}
\end{figure}

We display in figure~\ref{fig:energy} the results obtained for the Hubbard
model in the strongly attractive and spin asymmetric regime. For both studied
polarizations and cluster sizes, this preliminary numerical application of our
approach proves the convergence and stability of the averaged energy against
a long imaginary-time propagation. The bound statistical errors at any $\tau$
give good evidence that the phase problem, as well as the sampling of regions
where walkers are almost orthogonal to the trial state, are well mastered. We
finally discuss the quality of the approximate ground state resulting from the
use of biased weights (\ref{EQ:37}). For this purpose we map the obtained
energy onto the equivalent repulsive model. For the $6 \times 6$ cluster, the
estimated value at $\tau = 10t$  is $\underline{\tilde{E}}_G =-37.29(2) t$
that compares very favorably to the virtually exact value $E_G = -37.41(6) t$
\cite{Shi14}. The latter was obtained from QMC calculations with HF walkers in
the repulsive sector, starting from a constrained-path approximation that is
later on released. In \cite{Shi14}, the trial state consists of a large
superposition of Slater determinants. It yields a sizeable improvement on the
simple restricted-path approach, with one single HF wavefunction. Indeed, the
corresponding energy is $-36.68(7) t$. Our phaseless QMC calculations with HFB
walkers outperforms this standard value by nearly 2\%. To our knowledge, no
released constraint results for the $8 \times 8$ cluster are available, and we
therefore compare with variational Monte Carlo simulations. Using an extended
BCS-Gutzwiller wavefunction, Eichenberger and Baeriswyl found the variational
bound $E_G \leq -36.04 t$ \cite{Eic07}. With $\underline{\tilde{E}}_G =
-37.36(4) t$, our scheme yields a lower energy. These results are encouraging
and need to be confirmed by a detailed examination of the physical
content of the reconstructed wavefunctions through, e.g., the calculation of
correlation functions. They will be the purpose of a forthcoming publication.

\section{Summary and Perspectives}

Summarizing, we introduced in this work a QMC theoretical framework amenable
to the computation of an approximate ground state of strongly correlated
superconducting fermions. It 
relies on HFB wavefunctions that undergo a Brownian motion in
imaginary time. As compared to standard auxiliary-field QMC schemes, each
stochastic path can absorb fermion pair condensation that otherwise
would require a large superposition of HF realizations. The efficiency is also
improved by guiding the dynamics to generate walkers according to the
importance of their overlap with a trial wavefunction. A restricted-path
approximation is further implemented to prevent the development of an
infinite-variance problem by adequately sampling the directions almost
orthogonal to the trial state. Finally, the notorious phase problem is
managed through a fixed phase imposed to the overlap with the approximate
ground state reached at large imaginary time. Contrary to real-space QMC
methods, simulations can be performed by choosing any single-particle
basis. Any physical quantity can also be estimated by applying an extension of
Wick's theorem that we have formulated in terms of Pfaffians to avoid the
combinatorial complexity of standard expansions in products of binary
contractions.  

In condensed-matter physics, we expect our framework to help shedding new
light on the microscopic mechanisms leading to the formation of unconventional
Cooper pairs, such as the ones realized in the superconducting cuprates and
heavy fermion materials. Besides, the phaseless QMC approach with stochastic
HFB wavefunctions could unravel the degree of intertwining of order parameters
arising in systems exhibiting long wave-length modes. Another field of
application lies in synthetic quantum matter with ultra-cold atoms that can
emulate attractive Fermi systems.  In particular, the formalism is well suited
to the investigation of rotating superfluid Fermi gases in the strongly
interacting regime. Exotic pairing modes induced by artificial spin-orbit
couplings or in multicomponent gases could be addressed too.   

\ack

The authors acknowledge the financial support of the
French Agence Nationale de la Recherche (ANR), through
the program Investissements d'Avenir (ANR-10-LABX-09-01), LabEx EMC3, the
R\'egion Basse-Normandie, the R\'egion Normandie, and the Minist\`ere de la
Recherche.

\end{document}